\documentclass[12pt,a4paper]{article}
\usepackage[utf8]{inputenc}
\RequirePackage[l2tabu, orthodox]{nag}

\usepackage{jheppub}
\usepackage{amsthm,amssymb,amsmath,epic,float}
\usepackage{rotating,epsfig,indentfirst,array,varioref}
\usepackage{appendix,tikz,mathtools}
\usepackage{setspace}
\usepackage{amsmath}
\usepackage{mathtools}
\usepackage{ dsfont }
\usepackage{ wasysym }
\usepackage{braket}
\usepackage{amsfonts}
\usepackage{url}
\usepackage{amssymb}
\usepackage{hyperref}
\usepackage{ upgreek }
\usepackage{graphicx}
\usepackage{pst-solides3d}
\usepackage{cancel}

\usepackage{color}
\usepackage{tikz}
\usetikzlibrary{decorations.pathreplacing}
\usetikzlibrary{calc}
\usepackage{enumitem}
\usepackage{hyphenat}
\usetikzlibrary{positioning}

\usepackage{jheppub}
\makeatletter
\def\@fpheader{\vspace{-.1cm}}
\makeatother

\title{\boldmath Explicit construction of states in orbifolds of products of $N=2$ Superconformal ADE Minimal models}

\author[a]{Boris Eremin,}
\author[b,c]{Sergej Parkhomenko}

\affiliation[a]{Kharkevich Institute for Information Transmission Problems, 127994
Moscow, Russia}
\affiliation[b]{Landau Institute for Theoretical Physics, 142432 Chernogolovka, Russia}
\affiliation[c]{Moscow Institute of Physics and Technology, 141700 Dolgoprudny, Russia}

\emailAdd {Eremin.ba@phystech.edu, spark@itp.ac.ru}

\abstract{We generalize the explicit construction of fields in orbifolds of products of $N=(2,2)$ minimal models, developed by A. Belavin, V. Belavin and S. Parkhomenko to include minimal models with D and E-type modular invariants. It is shown that spectral flow twisting by the elements of admissible group $G_{\text{adm}}$, which is used in the construction of the orbifold, is consistent with the nondiagonal pairing of D and E-type minimal models. We obtain the complete set of fields of the orbifold from the mutual locality and other requirements of the conformal bootstrap. The collection of mutually local primary fields is labeled by the elements of dual group $G^{*}_{\text{adm}}$. The permutation of $G_{\text{adm}}$ and $G^*_{\text{adm}}$ is given by the mirror spectral flow  construction of the fields and maps the space of states of the original $G_{\text{adm}}$ orbifold onto the space of states of $G^*_{\text{adm}}$ orbifold.  We show that this transformation is by construction a mirror isomorphism of spaces of states. Thus, mirror isomorphism of states is built into the construction.
We illustrate our approach for the orbifolds of $\textbf{A}_{2}\textbf{E}_7^{3}$ model.}

\keywords{ADE Classification, Minimal model, Superconformal Field Theory, Mirror Symmetry, Conformal Bootstrap}

\begin{document} 
\maketitle
\flushbottom

\section{Introduction}
\label{sec:intro}

Conformal Field Theory models with $N=2$ supersymmetric extension of Virasoro algebra arise in the models of superstring compactification as degrees of freedom of compact sector \cite{BDFM:1988,CHSW:1985}. Gepner proposed that compactification of 6 out of 10 dimensions on Calabi-Yau manifold is equivalent to compactification on a unitary $N=2$ Superconformal field theory with central charge $c=9$. Moreover, by constructing certain orbifolds of the products of $N=(2,2)$ superconformal minimal models with the central charge $9$, he was able to show the equivalence of these orbifolds to $\sigma$-models on some CY-manifolds \cite{Gepner:1987vz,Gepner:1988}.

The approach of Gepner is based on the axioms of modular bootstrap, which implies the construction of a modular invariant partition function of characters of spectral flow twisted representations as a main ingredient of the construction.

In the papers \cite{BP,BBP,IIAB} we developed the explicit construction of fields in the orbifolds of products of $N=2$ minimal models, which is based on the conformal bootstrap approach as well as the spectral flow twisting. There it was implied that the minimal models used have a diagonal way of pairing of holomorphic and anti-holomorphic factors of the primary fields, i.e. type \textbf{A} minimal models were considered.

However, it is well known \cite{Cappelli:1987xt,Gepner:1988}, that the total list of $N=2$ superconformal minimal models includes also the models with pairing of type \textbf{D} and \textbf{E}. The problem we consider in this paper is the explicit construction of fields for the orbifolds of products that contain minimal models of type \textbf{D} and \textbf{E} as factors.

We begin with Section \ref{sec:2}, where we briefly discuss the spaces of states in models of types \textbf{A}, \textbf{D}, and \textbf{E}. At the end of the section, the $\textbf{E}_{7}$ model is considered as an example, and the total set of primary fields of the model is written out. 

In Section \ref{sec:3} the composite models, that is the products of $N=(2,2)$ superconformal minimal models with the diagonal $N=2$ Virasoro superalgebra of symmetries are considered. We analyze the group of discrete symmetries and define the admissible group $G_{\text{adm}}$ of orbifold. Then we represent our explicit construction of fields, which is based on the spectral flow twisting and mutual (qusi)-locality requirements of the fields of orbifold. 

In Section \ref{sec:4} we analyze the solutions of the mutual locality equations found in Section \ref{sec:3}. We use so-called mirror spectral flow twisting and require the conformal bootstrap axioms to hold. It allows us to define the dual admissible group $G_{\text{adm}}^*$.
We show that while the set of twisted sectors of the orbifold is given by the elements of $G_{\text{adm}}$, the set of mutually local primary fields of the orbifold is labeled by the elements of the group $G_{\text{adm}}^*$.
We show that the space of states of the original orbifold can be constructed using the dual group as the twisting sectors generation group. The permutation of these groups is nothing less but a mirror symmetry isomorphism of the spaces of states of the orbifold models. In other words, we show that mirror symmetry of states is built into the construction.
Moreover, the mirror symmetry is possible due to certain self-consistency of pairs of mutually dual groups with the spectrum of charges of primary fields in the composite model.

In Section \ref{sec:5} two mirror pairs of orbifolds of the composite model $\textbf{A}_{2}\textbf{E}_7^{3}$ are considered to illustrate our construction. That is, the composite model we are considering is the product of one minimal model with \textbf{A}-type modular invariant at level 1 and three minimal models with $\textbf{E}_7$-type modular invariant.  

The first pair of mirror models is given by the orbifolds by the minimal admissible group and its dual. We write out $(a,c)$ fields of these models and compare them with the De Rham cohomology of the corresponding Calabi-Yau manifolds. 

For the second pair of mirror models, the first admissible group is the one used by D. Gepner when constructing the 3-generation model. We find the dual group and construct the mirror orbifold. For both models we write out $(a,c)$ fields and compare them also with the De Rham cohomology of the corresponding Calabi-Yau manifolds. At the end of the section we briefly discuss the farther quotient of the second orbifold of these pair by the cyclic permutation group acting on the three $\textbf{E}_7$ factors
to obtain $(c,c)$ and $(a,c)$ fields in Gepner’s three-generation model.

We end the paper with the conclusion section.

\section{Models of $N=(2,2)$ SCFT and ADE classification}
\label{sec:2}


Recall that the unitary series of $N=(2,2)$ superconformal Minimal models $M_k$ are numbered with integer $k$ and have central charge 

\begin{equation}
    c=\frac{3k}{k+2}, \quad k=1,2,\dots
\end{equation}

The fields of the model can be distributed over certain products of minimal representations of the holomorphic and anti-holomorphic $N=2$ Virasoro superalgebra. For a given $k$ there exist finitely many representations of such kind. They are generated from primary states $\Phi_q^l$ with $l=0,\dots,k$ and $q=-l,-l+2,\dots,l$, which are the highest vectors of $N=2$ Virasoro superalgebra. In the $NS$ sector, the primary state is determined by the conditions
\begin{equation}
\label{eq:verma}
\begin{gathered}
 L_n \Phi^l_q=0, \ J_n\Phi^l_q=0, \ n\geq 1, \quad G_r^{\pm}\Phi^l_q=0, \ r\geq \frac{1}{2}
\\
L_0\Phi^l_q=\Delta_{l,q}\Phi^l_q, \  J_0 \Phi^l_q=Q_{l,q}\Phi^l_q.
\end{gathered}
\end{equation}
In the $R$ sector, the definition of the primary state is distinguished by the action of the fermionic generators of the $N=2$ Virasoro algebra:
\begin{equation}
    G_n^{\pm}\Phi^l_q=0,\ n\geq 1, \quad G_0^{+}\Phi^l_q=0.
\end{equation}

The $NS$ sector primary state $\Phi_q^l$ has a conformal dimension and $U(1)$ charge:

\begin{equation}
\label{eq:dimensionNS}
\Delta^{NS}_{l, q}=\frac{l(l+2)-q^2}{4(k+2)}, \quad Q^{NS}_{l, q}=\frac{q}{k+2} ,
\end{equation}
and in the $R$ sector the corresponding conformal weight and charge are
\begin{equation}
\label{eq:dimensionR}
\Delta_{l, q}^R=\frac{l(l+2)-(q-1)^2}{4(k+2)}+\frac{1}{8}, \quad Q_{l, q}^R=\frac{q-1}{k+2}+\frac{1}{2}.
\end{equation}

A key class of representations in the $NS$ sector are those generated by chiral-primary states, which are defined by an additional constraint:
\begin{equation}
    G^+_{-1/2}\Phi^l_q=0.
\end{equation}
Similarly, an anti-chiral primary state in the $NS$ sector must satisfy, in addition to (\ref{eq:verma}), the following condition:
\begin{equation}
    G^-_{-1/2}\Phi^l_q=0.
\end{equation}
These states are BPS, so the additional constraints they obey require $\Delta^{NS}=\frac{|Q^{NS}|}{2}$ \cite{Lerche:1989uy}.

Recall that in minimal model the chiral-primary and anti-chiral-primary states
\begin{equation}
\label{eq:chiral}
\Phi^l_c \equiv \Phi_{l}^{l}, \quad \Phi_a^l \equiv \Phi_{-l}^{l},
\end{equation}
appear when $q=\pm l$. 

In the models of $N=(2,2)$ SCFT, the primary fields are composed of the appropriately paired holomorphic and anti-holomorphic primary states.
In the modular bootstrap approach, the particular way of pairing can be read off from the partition function. As it was found by D.Gepner \cite{Gepner:1988} the most general modular invariant partition function of the minimal model takes the form
\begin{equation}
\label{eq:partfMIN}
Z(\tau)=    \frac{1}{2} \sum_{l,\bar{l}=0}^k \sum_{\substack{q,\bar{q}=-(k+1);  \\ l+q+s=0, \ \bar{l}+\bar{q}+\bar{s}=0 \ \text{mod} \ 2}}^{k+2}\sum_{\substack{s,\bar{s}=0}}^{3} N_{l,\bar{l}}M_{q,\bar{q}}L_{s,\bar{s}} \chi_{q,s}^{l}(\tau)\bar{\chi}_{\bar{q},\bar{s}}^{\bar{l}}(\bar{\tau}),
\end{equation}
where $\chi_{q,s}^{l}$, $s=0,1,2,3$ are the splited characters of the highest weight representation $\mathcal{H}^l_{q}$ of $N=2$ Virasoro superalgebra, which are calculated over the subspaces with a fixed fermion number. More precisely, $s=0,1$ give the characters calculated over the subspaces with even number of fermionic generators of $G^{\pm}$  in $NS$ and $R$ sectors correspondingly, while $s=2,3$ give the characters calculated over the subspaces with odd number of fermionic generators of $G^{\pm}$ in $NS$ and $R$ sectors.
The matrices $N_{l,\bar{l}},M_{q,\bar{q}},L_{s,\bar{s}}$ provide modular invariance of the partition function (see also \cite{Gepner:Parafermion}, \cite{Gepner:Lectures}). 

Among them, the most important matrix $N_{l,\bar{l}}$ was found by Cappelli, Itzykson, and Zuber \cite{Cappelli:1987xt} when they addressed the problem of modular invariance for the conformal field theory models with affine $\widehat{s l}(2)$ algebra of symmetries. 

They found the list

\begin{equation}
\label{eq:ADE}
\begin{aligned}
& \left(\textbf{A}_{k+1}\right), k \geq 1: \sum_{l=0}^k \chi_l \bar{\chi}_l \\
& \left(\textbf{D}_{2 j+2}\right) , k=4 j: \sum_{l=0}^{j-1}\left|\chi_{2 l}+\chi_{k-2 l}\right|^2+2 \chi_{k / 2} \bar{\chi}_{k / 2} 
, \\
& \left(\textbf{D}_{2 j+1}\right), k=4 j-2: \sum_{l=0}^{k/2}\left|\chi_{2 l}\right|^2+\sum_{l=0}^{k/2-1} \chi_{2 l +1} \bar{\chi}_{k-2l-1} 
, \\
& \left(\textbf{E}_6\right), k=10:\left|\chi_0+\chi_6\right|^2+\left|\chi_3+\chi_7\right|^2+\left|\chi_4+\chi_{10}\right|^2, \\
& \left(\textbf{E}_7\right), k=16: \left|\chi_0+\chi_{16}\right|^2+\left|\chi_4+\chi_{12}\right|^2+\left|\chi_6+\chi_{10}\right|^2+\left|\chi_8\right|^2 + \left(\chi_2+\chi_{14}\right) \bar{\chi}_8+\chi_8\left(\bar{\chi}_2+\bar{\chi}_{14}\right) , \\
& \left(\textbf{E}_8\right), k=28:\left|\chi_0+\chi_{10}+\chi_{18}+\chi_{28}\right|^2 +\left|\chi_6+\chi_{12}+\chi_{16}+\chi_{22}\right|^2, \\
\end{aligned}
\end{equation}
which includes the simplest diagonal invariant of type $\textbf{A}_{k+1}$ for all integer levels $k\geq 1$. For even $k$, two series of off-diagonal invariants of type $\textbf{D}$ arise: $\textbf{D}_{2 j+2}$ for $k=4j$, and $\textbf{D}_{2 j+1}$ for $k=4j-2$. Besides these infinite series, there are also three exceptional cases: $\textbf{E}_6, \textbf{E}_7,$ and $\textbf{E}_8$ for $k=10,16,$ and 28 respectively. The alternative proof of the classification (\ref{eq:ADE}) can be found in \cite{Gannon:1999cp}.

In the very nice paper \cite{Gannon:1996hp} the more general list of modular invariants for the $N=(2,2)$ superconformal minimal models has been found. But in this paper we restrict ourself to the models whose partition functions are given by (\ref{eq:partfMIN}), where the matrix $N_{l,\bar{l}}$ is given by (\ref{eq:ADE}) and the matrices $M$, $L$ are identity matrices.

It is worth noting that the expression for the partition function in (\ref{eq:partfMIN}) is, in a sense, formal. In it
the parameter $q$ goes through all possible values $q=-(k+1),\dots,(k+2)$ and depends on $l$ through the parity condition: $ l+q=0 \  \text{mod} \ 2$ in $NS$ sector, and $l+q=1 \  \text{mod} \ 2$ in $R$ sector. However, the minimal representations of the $N=2$ Super-Virasoro algebra are defined only for $q=-l,-l+2,\dots,l$. Consequently, the corresponding highest vector $\Phi^l_q$ of the representation $\mathcal{H}^l_{q,s}$ with $|q|>l$ does not belong to the Kac table of the minimal model $M_{k}$ and their conformal dimension and $U(1)$ charge can not be computed using (\ref{eq:dimensionNS}). Despite this, the symbolic form of (\ref{eq:partfMIN}) facilitates the proof of modular invariance, since the indices $(l,\bar{l})$, $(q,\bar{q})$, and $(s,\bar{s})$ in the product $N_{l,\bar{l}} M_{q,\bar{q}} L_{s,\bar{s}}$ decouple under modular transformations \cite{Gepner:1988}.

To clarify the interpretation of the remaining terms in the partition function, we can rely on the isomorphisms of $N=2$ Virasoro superalgebra representations \cite{Gepner:1987vz}, \cite{Gepner:Lectures}, \cite{Feigin}, established rigorously in \cite{FSST} as the isomorphisms of twisted representations:

\begin{equation}
    \label{eq:identification0}
    \mathcal{H}^l_q\simeq \mathcal{H}^{k-l}_{q+(k+2)}, \quad \mathcal{H}^l_q\simeq \mathcal{H}^{l}_{q+2(k+2)}.
\end{equation}
These isomorphisms allow us to clarify the meaning of the terms appearing in (\ref{eq:partfMIN}) that correspond to the states $\Phi^l_q$ with $|q|>l$. They should be understood via the identifications
\begin{equation}
    \label{eq:identification}
    \Phi_{q}^l\simeq \Phi_{q+(k+2)}^{k-l}, \quad\Phi_{q}^l\simeq \Phi_{q+2(k+2)}^l.
\end{equation}
Below in our article, especially in section \ref{sec:5} with examples, the states $\Phi^l_q$ with $|q|>l$ may arise. They should be understood as being identified with $\Phi^{k-l}_{q\pm(k+2)}$. For now, the identifications (\ref{eq:identification0}), (\ref{eq:identification}) are more of an enumerative  nature. Below, in section \ref{sec:3}, we will explain the exact meaning of these identifications.

Isomorphisms of representations (\ref{eq:identification0}) provide equality of corresponding characters
\begin{equation}
    \chi_{q,s}^l(\tau)=\chi_{q+(k+2),s+2}^{k-l}(\tau), \quad \chi_{q}^l(\tau)= \chi_{q+2(k+2),s+4}^l(\tau).
\end{equation}
Using these it is easy to explain the coefficient $\frac{1}{2}$ in the partition function (\ref{eq:partfMIN}). The point is that the parameters $(l,q,s)$ and $(\bar{l},\bar{q},\bar{s})$ run over all possible values, so there are terms with characters $\chi_{q,s}^{l}(\tau)$ and $\chi_{q+(k+2),s+2}^{k-l}(\tau)$. Due to their equality, each term appears twice. However, this is not related to multiplicity, but arises due to the symbolic notation of the partition function (\ref{eq:partfMIN}), as we explained above. The only exception is the \textbf{D} series for $k=4j$. In it the multiplicity of characters with $l=\bar{l}=\frac{k}{2}$ is 2. This amazing fact has been discussed, for example, in the paper \cite{Fuchs:1991vu}.

In this way we can describe the spectrum in $N=(2,2)$ minimal models for a given $k$. It contains the mutually local primary fields of the form:
\begin{equation}
\label{eq:FIELDS}
\begin{aligned}
& \Psi_{q}^{l,\bar{l}}(z, \bar{z})=\Phi_{ q}^{l}(z) \bar{\Phi}_{ q}^{\bar{l}}(\bar{z}), 
\end{aligned}
\end{equation}
where the right and left indexes $l$ and $\bar{l}$ are correlated according to \textit{ADE classification} (\ref{eq:ADE}). 

Of course the holomorphic factorization we depicted in (\ref{eq:FIELDS}) should not be taken literally. The summation of quantum group indices, which ensures mutual locality of fields within correlation functions, is certainly implied and will be implied in subsequent similar expressions.
But these details are not very important for the discussion in the paper. So, in what follows we will use rather symbolic expressions like (\ref{eq:FIELDS}) instead of explicit formulas. 

Taking into account the identification (\ref{eq:identification}) we can assume that in $NS$ sector the charge $q$ runs:
\begin{equation}
\label{eq:q}
    q=-(k+1), \dots , (k+2), \quad q+l=0 \ (\text{mod}   \ 2),
\end{equation}
and in $R$ sector: 
\begin{equation}
\label{eq:qR}
    q=-(k+1), \dots , (k+2), \quad q+l=1 \ (\text{mod}   \ 2).
\end{equation}

Actually, some of the ADE invariants can be decomposed \cite{Fuchs:1991vu} as: $4_D=1_A\otimes 1_A$, $10_E=1_A\otimes 2_A$, and $28_E=1_A\otimes 3_A$.

\subsection{Example: spectrum of $\textbf{E}_7$ minimal model}

Let us show how one can find the spectrum of the minimal model, defined by the modular invariant (\ref{eq:partfMIN}) using the example of the invariant $\textbf{E}_7$ with $k=16$. The partition function of the corresponding $\widehat{s l}(2)$ WZW model is

\begin{equation}
\label{eq:e7}
Z_{E_7}=\left|\chi_0+\chi_{16}\right|^2+\left|\chi_4+\chi_{12}\right|^2+\left|\chi_6+\chi_{10}\right|^2+\left|\chi_8\right|^2+\left(\chi_2+\chi_{14}\right) \bar{\chi}_8+\chi_8\left(\bar{\chi}_2+\bar{\chi}_{14}\right).
\end{equation}

So, the matrix $N_{l,\bar{l}}$ of the Minimal model can be read off from here. Since the matrices $M$, $L$ are identity matrices, we can easily write out the list of $NS$ primary fields of the Minimal model $\textbf{E}_7$ taking into account the identifications (\ref{eq:identification}). This set is shown in table \ref{tabular:E7}.

The first column contains the term from the partition function, which determines the corresponding matrix elements $N_{l,\bar{l}}$ and corresponds to the fields in the second column. Since these are fields from the $NS$ sector, $q+l$ is even for all fields in the table.
Note that some fields have left and right states with different charges: $\bar{q}=q\pm(k+2)$. It arises from the identification rules (\ref{eq:identification}) when we bring the labels $q, \bar{q}$ in to the domain  $|q|\leq l, |\bar{q}|\leq\bar{l}$. It is also straightforward to list the primary fields of the $R$-sector as well.

\section{Orbifolds of composite models}
\label{sec:3}

Taking the products of $N=(2,2)$ minimal models with total central charge $9$ 
\begin{equation}
M_{\vec{k}}=\prod_{i=1}^{r}M_{k_{i}}, \quad \sum_{i=1}^r \frac{3k_{i}}{k_{i}+2}=9,
\label{eq:prodmin}
\end{equation} 
we obtain so-called composite models, where each factor $M_{k_{i}}$ is a minimal model of ADE-type. Certain orbifolds of these models are closely related with the
$\sigma$-models on CY manifolds \cite{Gepner:1987vz,Gepner:1988} at some points of moduli space. The more precise connection of the composite model orbifold with nondiagonal modular invariant and the Calabi-Yau hypersurface was established in \cite{Fuchs:1989pt, Lynker}.

The $N=(2,2)$ Virasoro superalgebra of the composite model $M_{\vec{k}}$ as well as its orbifold, is defined as the diagonal subalgebra in the tensor product of models:
\begin{equation}
L_{orb,n}=\sum_{i}L_{(i),n}, \ J_{orb,n}=\sum_{i}J_{(i),n},
\
G^{\pm}_{orb,r}=\sum_{i}G^{\pm}_{(i),r}.
\label{eq:Vir}
\end{equation}
The action of this algebra is correctly defined only on the product of $NS$ representations  or on the product of  $R$  representations of minimal models $M_{k_{i}}$. Therefore, we can form $NS$ or $R$ primary fields  in the composite model $M_{\vec{k}}$ by taking only products of primary fields from each minimal model $M_{k_{i}}$  belonging to  the same ($NS$ or $R$) sectors:
\begin{equation}
\label{eq:composit}
\begin{aligned}
\Psi^{\vec{l},\bar{\vec{l}}}_{\vec{q}}(z, \bar{z})=\prod_i \Psi_{q_i}^{l_i,\bar{l}_i}(z, \bar{z}).
\end{aligned}
\end{equation}
(recall that we use the same notation for the $NS$ fields and $R$ fields as long as it does not cause confusion), where the left and right numbers $l_i, \bar{l}_i$ are connected according to the ADE classification described above. 

The fields above, of course, do not exhaust the entire list of primary fields of the composite model. However, they are especially important for us since all other fields in the composite model can be obtained from them 
by applying the operators of the $N = (2, 2)$ Virasoro superalgebras
of $M_{k_{i}}$ models.

\subsection{Group of symmetry of the composite model}
\label{sec:group}

The orbifold is defined by the admissible group $G_{\text{adm}}$ \cite{BH,Kra}. It is a subgroup of the group of total abelian symmetry of the composite model $M_{\vec{k}}$: 
\begin{equation}
    G_{\text{tot}}=\prod_{i=1}^rG_i= 
    \left\{\prod_i \hat{g}_i^{w_i}, w_i \in \mathbb{Z}, \hat{g}_i=\exp \left(i 2 \pi J_{(i), 0}\right)\right\},
    \label{eq:Gtot}
\end{equation}
where $G_{i}$ is the discrete symmetry group of the individual model $M_{k_{i}}$.
The $G_{\text{tot}}$ depends on the modular invariant (\ref{eq:ADE}) in each copy of $M_{k_i}$. 

In the $\textbf{D}_{2j+2}$ type series for $k=4j$, as well as in $\textbf{E}_7$ and $\textbf{E}_8$ the parameters $l_i,\bar{l}_i$ run through only even values. Then the charges $q_{i}$ of primary state are also only even. It follows that for the corresponding element $\hat{g}_i^{w_i}\in G_i$, the exponent $w_i$ is defined modulo $(k_i+2)/2$, then $G_i\simeq\mathbb{Z}_{(k_i+2)/2}$. For modular invariants of type $\textbf{A}_{k_i+1}$ and $\textbf{D}_{2j+1}$ with $k_i=4j-2$ as well as for the exceptional $\textbf{E}_{6}$, there are no restrictions on the parity of the charges. Then the corresponding group is $G_i\simeq \mathbb{Z}_{k_i+2}$. 
The case of diagonal invariants \cite{BBP} 
has a group $G_{\text{tot}}=\prod_{i=1}^r \mathbb{Z}_{k_i+2}$.

Below in the text, it will be useful to use additive version of the group $G_{tot}$ labeling its elements by the vectors $\vec{w}=(w_{1},...,w_{r})$
and regard the elements of the group $w_i\in\mathbb{Z}_{(k_i+2)/2} $ as \textit{even} elements  in $\mathbb{Z}_{k_i+2}$. We will explain this restriction in more detail in section \ref{sec:4}.

The \textit{admissible group} is defined as any subgroup of $G_{\text{tot}}$, satisfying

\begin{equation}
\label{eq:adm}
G_{\text{adm}}=\left\{ \vec{w}  =(w_1,\dots,w_r) \bigg| \quad w_i\in G_i, \quad \sum_i^r \frac{w_i}{k_i+2} \in \mathbb{Z} \right\}\subset G_{tot}.
\end{equation}
As will be seen (see section \ref{sec:5}), the constraint
\begin{equation}
\sum_{i}^{r}\frac{w_{i}}{k_{i}+2}\in\mathbb{Z},
\label{eq:CYcondition}
\end{equation}
is nothing else but the requirement that the $(c,c)$ primaries, charged as $(3,0)$ and $(0,3)$ (which, in addition, are the holomorphic and anti-holomorphic spin-3/2 currents respectively)
are among mutually local fields of the orbifold model. Thus, the requirement (\ref{eq:CYcondition}) is necessary since these currents must be identified with a nowhere-vanishing holomorphic $(3,0)$ and anti-holomorphic $(0,3)$ forms on the Calabi-Yau manifold 
\cite{EOTY}. We construct these currents explicitly in section \ref{sec:5}.

Notice that in case of an odd number of models $r$, as well as when choosing a modular invariant of type $\textbf{A}_{k_i+1}$, $\textbf{D}_{2j+1}$, or $\textbf{E}_{6}$ in each $M_{k_i}$, there exists the element $\vec{w}=(1,\dots,1)\in G_{\text{adm}}$. For even $r$ or if there is the invariant of type $\textbf{D}_{2j+2}$, $\textbf{E}_7$, or $\textbf{E}_8$ in some $M_{k_i}$ (then the corresponding $w_i$ is even and defined modulo $k_i+2$ as we explained above), the simplest solution of the admissibility condition (\ref{eq:CYcondition}) is $\vec{w}=(2,\dots,2)$.

\subsection{Spectral Flow twisting and mutual locality of fields}
\label{sec:spectralflow}

As the admissible group is fixed, we can use it to extend the initial space of fields of the composite model by adding the twisted fields. This is the first step of our construction of fields in the orbifold model. In the second step, we extract the orbifold fields as a subspace that contains only mutually (quasi-)local fields. The quasi-locality appears because we want to include also fields from the $R$-sector, where the $\pm$ monodromy of the correlation functions is allowed. Thus, our procedure is completely similar to the case of composite models with the \textbf{A}-type modular invariants \cite{BBP, Parkhomenko:2022kju} (and is a special case of the more general approach discussed in \cite{ShYan})
.

In order to implement the first step, we use the spectral flow realization of primary states \cite{BP,BBP, IIAB}. It goes as follows. 

Recall the spectral flow automorphism of $N=2$ Super-Virasoro algebra \cite{SS}:
\begin{equation}
\begin{aligned}
& \tilde{G}_r^{ \pm}=U^{-t} G_r^{ \pm} U^t=G_{r \pm t}^{ \pm}, \\
& \tilde{J}_n=U^{-t} J_n U^t=J_n+\frac{c}{3} t \delta_{n, 0}, \\
& \tilde{L}_n=U^{-t} L_n U^t=L_n+t J_n+\frac{c}{6} t^2 \delta_{n, 0} .
\end{aligned}
\end{equation}
For values $t \in \mathbb{Z}+\frac{1}{2}$ the spectral flow maps $NS$ and $R$ sectors into each other, and for $t \in \mathbb{Z}$ it acts inside $NS$ or $R$ sectors. 

The extension of the spectral flow automorphism above to the $N=2$ Virasoro superalgebra representations \cite{Lerche:1989uy} generates so-called twisted representations \cite{FSST}. In particular, the initial highest weight vector $\Phi^{l}_{q}$ becomes the twisted ones after the spectral flow operator $U^{t}$ action. At the same time, exactly the characters of these twisted representations appear in Gepner's modular invariant partition function of superstring compactification. It raises the question of whether we can find the standard highest weight vector among the states of the twisted representations? 

The answer is given by the couple of formulas below \cite{BBP}. The state
\begin{equation}
V^l_{t}=(UG^{-}_{-\frac{1}{2}})^{t}\Phi^{l}_{c}, \quad 0\leq t\leq l.
\label{eq:Prim1}
\end{equation}
gives spectral flow realization of the $NS$ primary state $\Phi^{l}_{q}$ of the individual minimal model, where $q=l-2t$. In other words, the expression (\ref{eq:Prim1}) gives the construction of the primary state $\Phi^{l}_{q}$ as a specific descendant in the chiral-primary representation with the chiral-primary state $\Phi^{l}_{c}$, which is twisted by $U^{t}$, where $0\leq t\leq l$.

For the purposes of the orbifold construction, we need to extend this formula. 
Namely, the state
\begin{equation}
V^l_{t}=(UG^{-}_{-\frac{1}{2}})^{t-l-1}U(UG^{-}_{-\frac{1}{2}})^{l}\Phi^{l}_{c}, \quad l+1\leq t\leq k+1
\label{eq:Prim2}
\end{equation}
gives the spectral flow realization of the $NS$ primary state $\Phi^{\tilde{l}}_{\tilde{q}}$, where $\tilde{l}=k-l$, $\tilde{q}=k+2+l-2t$. The expression (\ref{eq:Prim2}) gives the construction of the primary state $\Phi^{\tilde{l}}_{\tilde{q}}$ as a specific descendant in the chiral-primary representation with the chiral-primary state $\Phi^{l}_{c}$, which is twisted by $U^{t}$, where $l+1\leq t\leq k+1$.

It is clear that the $R$ primary state is given by $U^{\frac{1}{2}}$ twisting of the $NS$ primary fields above. The validity of formulas (\ref{eq:Prim1}), (\ref{eq:Prim2}) follows from the structure of singular vectors of unitary, integrable representations of the $N=2$ super-Virasoro algebra \cite{FSST}, \cite{Feigin}. Below we generalize these formulas to construct the primary fields in the orbifold model.

Notice that for $t=k+2$ the state
\begin{equation}
V^l_{k+2}=U(UG^{-}_{-\frac{1}{2}})^{k-l}U(UG^{-}_{-\frac{1}{2}})^{l}\Phi^{l}_{c}
\label{eq:Prim3}
\end{equation}
gives the spectral flow realization of the primary state which can be identified with the original chiral primary one $\Phi^l_c$. 
From this perspective, the action of the spectral flow operator on the representations has a ''periodicity'' condition
\begin{equation}
\label{eq:pereodicity}
    U^{k+2}\simeq1. 
\end{equation}
As it is clear from (\ref{eq:Prim3}) this relation should not be interpreted literally but as an isomorphism between representations generated by highest vectors (\ref{eq:Prim1})-(\ref{eq:Prim2}) at $t=0$ and $t=k+2$.
Consequently, the parameter $t$ is defined modulo $k+2$. 

The previously mentioned identification of states (\ref{eq:identification}), should also be understood within the framework of the relations (\ref{eq:Prim1}),(\ref{eq:Prim2}). Indeed, while (\ref{eq:Prim1}) provides the spectral flow realization of the primary state $\Phi^{l}_{q}$, where $q=l-2t$ and hence, $|q|\leq l$, one can consider the expression (\ref{eq:Prim2}) as giving spectral flow realization of some primary state $\Phi^{l}_{q}$, where again $q=l-2t$, but this time $l\leq |q|$. From the other hand, this state is nothing else but the primary state $\Phi^{k-l}_{q+k+2}$, where $|q+k+2|\leq k-l$. It yields the first identification from (\ref{eq:identification}). This implies also isomorphisms of representations (\ref{eq:identification0}). Given (\ref{eq:Prim3}), the second identification from (\ref{eq:identification}) should be understood similarly.


We also note that it is possible to extend the domain of the spectral flow parameter $t$ to the full set of integers and obtain an infinite set of realizations of the primary states by appropriately generalizing the expressions (\ref{eq:Prim2}). This amazing fact has already been implied in \cite{FST} and is certainly worth being considered further. However, we leave this issue beyond the scope of the present paper.

In terms of spectral flow realization of the primary states, the primary field $\Phi_{q}^l(z)\bar{\Phi}^{\bar{l}}_q(\bar{z})$ from the minimal model $M_k$ with explicit ADE invariant can be realized as follows
\begin{equation}
    \label{eq:primaryspectral}
    \Psi^{l,\bar{l}}_{t,\bar{t}}(z,\bar{z})=V_{t}^l(z)\bar{V}^{\bar{l}}_{\bar{t}}(\bar{z}).
\end{equation}
Note that, unlike the fields from the $\textbf{A}_{k+1}$ series, the spectral flow parameter $t$ and $\bar{t}$ for the left- and right-moving sectors may differ, although their charges coincide modulo the identification (\ref{eq:identification}): $q=l-2t=\bar{q}=\bar{l}-2\bar{t}$. This occurs due to the possible difference in the parameters $l$ and $\bar{l}$.

Now we use the spectral flow realizations (\ref{eq:Prim1}), (\ref{eq:Prim2}), (\ref{eq:primaryspectral}) to build twisted states from the states of the composite model for each element of the admissible group $\vec{w} \in G_{\text{adm}}$.

In $NS$ sector the twisted primary field is given by
\begin{equation}
\label{eq:twistfield}
\Psi_{\vec{t}+\vec{w},\vec{\bar{t}}}^{\vec{l},\vec{\bar{l}}}(z, \bar{z})=V_{\vec{t}+\vec{w}}^{\vec{l}}(z) \bar{V}_{\vec{\bar{t}}}^{\vec{\bar{l}}}(\bar{z}),
\end{equation}
where
\begin{equation}
\label{eq:rightstate}
\bar{V}_{\vec{\bar{t}}}^{\vec{\bar{l}}}(\bar{z})=\prod_i^r \bar{V}_{ \bar{t}_i}^{\bar{l}_i}(\bar{z}),\quad \bar{V}_{ \bar{t}_i}^{\bar{l}_i}(z)=\left(U G_{-\frac{1}{2}}^{-}\right)_i^{\bar{t}_i} \bar{\Phi}^{\bar{l}_i}_c(z), \quad  0 \leq \bar{t}_i \leq \bar{l}_i
\end{equation}
and
\begin{equation}
\label{eq:leftstate}
V_{\vec{t}+\vec{w}}^{\vec{l}}(z)=\prod_{i=1}^r V_{t_i+w_i}^{l_i}(z),
\end{equation}
where
\begin{equation}
\label{eq:leftstate1}
V_{t_i+w_i}^{l_i}(z)= \begin{cases}\left(U G_{-\frac{1}{2}}^{-}\right)_i^{t_i+w_i} \Phi^{l_i}_c(z), & \text { if } \quad 0 \leq t_i+w_i \leq l_i, \\ \left(U G_{-\frac{1}{2}}^{-}\right)_i^{t_i+w_i-l_i-1} U_i\left(U G_{-\frac{1}{2}}^{-}\right)_i^{l_i} \Phi^{l_i}_c(z), & \text { if } \quad l_i+1 \leq t_i+w_i \leq k_i+2.\end{cases}
\end{equation}
Note that for $\vec{w}=0$, the expressions (\ref{eq:twistfield})-(\ref{eq:leftstate1}) give the spectral flow construction of primary fields (\ref{eq:composit}) from $NS$ sector of the original composite model, where $\vec{q}=\vec{l}-2\vec{t}=\vec{\bar{l}}-2\vec{\bar{t}}$.

The second step is to satisfy mutual locality of the fields (\ref{eq:twistfield}). In this way we can determine the spectrum of the orbifold model.

Consider two twisted fields:
\begin{equation}
\label{eq:couple}
\begin{aligned}
& \Psi_{\vec{t}_1+\vec{w}_1,\vec{\bar{t}}_1}^{\vec{l}_1,\vec{\bar{l}}_1}(z, \bar{z}), \quad \Psi_{ \vec{t}_2+\vec{w}_2,\vec{\bar{t}}_2}^{\vec{l}_2, \vec{\bar{l}}_2}(0,0), \\
& \vec{w}_1, \vec{w}_2 \in G_{\text{adm}}
\end{aligned}
\end{equation}
and compute the phase factor $\exp(i\theta)$ arising from the monodromy $z\rightarrow e^{i\theta}z$. To simplify the calculation of the phase factor one can use free bosonic field $\phi_i(z)$ representation for the spectral flow operator and $U(1)$ current $J_{(i)}(z)$:
\begin{equation}
\begin{gathered}
J_{(i)}(z)=i \sqrt{\frac{k_i}{k_i+2}} \partial \phi_i(z), \quad U_i(z)=\exp \left(i \sqrt{\frac{k_i}{k_i+2}} \phi_i(z)\right),
\end{gathered}
\end{equation}
where it is implied that
\begin{equation}
    \phi_{i}(z)\phi_{j}(0)=-\delta_{ij}\log(z)+....
\end{equation}

Then the primary states of the minimal model factors of the composite model, as is well known, can be represented as \cite{ZF}

\begin{equation}
\label{eq:paraferm}
\Phi_{q_i}^{l_i}(z)=\exp \left(i \frac{q_i}
{\sqrt{k_i\left(k_i+2\right)}} \phi_i(z)\right) \hat{\Phi}_{q_i}^{l_i}(z),
\end{equation}
where the $\hat{\Phi}_{q_i}^{l_i}(z)$ is uncharged with respect to current $J_i(z)$. The complete primary field of the model $M_{k_i}$ is an appropriate product of the factor above and
an antiholomorphic factor that can be represented similarly (as we have already mentioned, this product must include quantum group indices summation).

We rely on the mutual locality of the fields of the composite model. 
As a result, in order to find $\exp(i\theta)$ we need to compute the phase factor of the operator product:

\begin{multline}
\exp \left[i \sum_i\left(\sqrt{\frac{k_i}{k_i+2}} w_{1 i}+\frac{q_{1 i}}{\sqrt{k_i\left(k_i+2\right)}}\right) \phi_i(z)\right]\times \\ \times \exp \left[i \sum_j\left(\sqrt{\frac{k_j}{k_j+2}} w_{2 j}+\frac{q_{2 j}}{\sqrt{k_j\left(k_j+2\right)}}\right) \phi_j(0)\right],
\end{multline}
where $q_{1,2 i}=l_{1,2 i}-2t_{1,2 i}$.
Applying the well-known result for exponents of free fields:

\begin{equation}
    :\exp \left(i\sum_i \alpha_i \phi_i(z) \right) :: \exp \left(i\sum_j \gamma_j \phi_j(0) \right): = z^{\sum_i \alpha_i\gamma_i}:\exp\left(i\sum_j\alpha_j\phi_j(z)+\gamma_j\phi_j(0) \right):
\end{equation}
we find the phase factor easily and obtain the mutual locality equation:
\begin{equation}
\label{eq:mutual}
    \sum_{i=1}^r \frac{w_{1i}\left(q_{2i}-w_{2i} \right)+w_{2i}\left(q_{1i}-w_{1i} \right)}{k_i+2} \in \mathbb{Z}.
\end{equation}

There is another way to get this equation which doesn't use the representation (\ref{eq:paraferm}). To this end one can consider the operator product expansion of the fields (\ref{eq:couple}) and demand the corresponding monodromy vanishes (see Appendix in \cite{IIAB}).

The equation (\ref{eq:mutual}) is completely analogous to the equation obtained for composite models of type $\textbf{A}$ in \cite{BBP, Parkhomenko:2022kju}.
Thus, the off-diagonality of the fields (\ref{eq:twistfield}) with respect to the indices $\vec{l},\vec{\bar{l}}$, arising for $\textbf{D}$ and $\textbf{E}$ invariants, affects the mutual locality condition only through the allowed set of vectors $\vec{q}$ which appear in the twisted sector $\vec{w}$.

The twisted field in the $R$ sector can be obtained by applying the spectral flow:
\begin{equation}
\begin{aligned}
\prod_i U_i^{\frac{1}{2}} \bar{U}_i^{\frac{1}{2}} \Psi_{t_i+w_i,\bar{t}_i}^{l_i,\bar{l}_i}(z, \bar{z}). 
\end{aligned}
\end{equation}

Thus, by solving the mutual locality equations (\ref{eq:mutual}) with respect to $\vec{q}$ in each twisted sector $\vec{w}$ and taking into account descendants, we can explicitly determine all fields of the orbifold model. It is easy to see that in this way we have obtained a set of (quasi-)local fields that also satisfy the other axioms of the conformal bootstrap. In other words, we obtain a model of a $N=(2,2)$ superconformal field theory.

\section{Mutual locality and Mirror model}
\label{sec:4}

The solutions to equation (\ref{eq:mutual}) can be investigated by analyzing the construction of the spectral flow of the dual orbifold \cite{Parkhomenko:2024mxq,Parkhomenko:2022kju} (whose notion will become clear soon). However, the remarkable fact is that the field space of the mirror orbifold arose automatically as soon as we constructed the state space of the original orbifold.

To verify this, let us consider the mirror version of (\ref{eq:twistfield}), (\ref{eq:mutual}) \cite{Parkhomenko:2022kju}. It uses so-called mirror spectral flow twisting, which starts with the anti-chiral primary field 
so the mirror realization of a Minimal model primary states is given by
\begin{equation}
\label{eq:mirrorspectral1}
\tilde{V}^l_{t}=\left(U^{-1} G_{-\frac{1}{2}}^{+}\right)^{l-t} \Phi^l_a, \quad 0 \leq t \leq l,
\end{equation}
\begin{equation}
\label{eq:mirrorspectral2}
\tilde{V}^l_{t}=\left(U^{-1} G_{-\frac{1}{2}}^{+}\right)^{k+1-t} U^{-1}\left(U^{-1} G_{-\frac{1}{2}}^{+}\right)^l \Phi^l_a, \quad l+1 \leq t \leq k+1.
\end{equation}

It allows one to get a mirror realization of the $NS$ primary fields (\ref{eq:twistfield}) in the form $\tilde{\Psi}^{\vec{l},\vec{\bar{l}}}_{\vec{t}+\vec{w},\vec{\bar{t}}}(z,\bar{z})=\tilde{V}_{\vec{t}+\vec{w}}^{\vec{l}}(z)\bar{V}_{\vec{\bar{t}}}^{\vec{l}}(\bar{z})$, where 
\begin{equation}
\begin{aligned}
& \tilde{V}_{\vec{t}+\vec{w}}^{\vec{l}}(z)=\prod_{i=1}^r \tilde{V}_{t_i+w_i}^{l_i}(z), \\
& \tilde{V}_{t_i+w_i}^{l_i}(z)=\left\{\begin{array}{l}
\left(U^{-1} G_{-\frac{1}{2}}^{+}\right)_i^{l_i-t_i-w_i} \Phi^{l_i}_a(z), \quad \text { if } \quad 0 \leq t_i+w_i \leq l_i, \\
\left(U^{-1} G_{-\frac{1}{2}}^{+}\right)_i^{k_i+1-t_i-w_i} U_i^{-1}\left(U^{-1} G_{-\frac{1}{2}}^{+}\right)_i^{l_i} \Phi^{l_i}_a(z), \quad \text { if } \quad l_i+1 \leq t_i+w_i \leq k_i+1 .
\end{array}\right.
\end{aligned}
\end{equation}
And $\bar{V}_{\vec{\bar{t}}}^{\vec{l}}(\bar{z})$ remains defined as in (\ref{eq:rightstate}).

Making the involution
\begin{equation}
G^{ \pm}(z) \rightarrow G^{\mp}(z), J(z) \rightarrow-J(z), U(z) \rightarrow U^{-1}(z), T(z) \rightarrow T(z),
\end{equation}
for the generators of holomorphic Virasoro superalgebra, the field $\tilde{\Psi}^{\vec{l},\vec{\bar{l}}}_{\vec{t}+\vec{w},\vec{\bar{t}}}(z,\bar{z})$ takes the old form
\begin{equation}
\begin{aligned}
& \Psi^{\vec{l},\vec{\bar{l}}}_{\vec{t}+\vec{w}^*,\vec{\bar{t}}}(z,\bar{z})=V_{\vec{t}+\vec{w}^*}^{\vec{l}}(z)\bar{V}_{\vec{\bar{t}}}^{\vec{l}}(\bar{z}), \\
& V_{\vec{t}+\vec{w}^*}^{\vec{l}}(z)=\prod_{i=1}^r V_{t_i+w_i^*}^{l_i}(z), \\
& V_{t_i+w_i^*}^{l_i}(z)=\left\{\begin{array}{l}
\left(U G_{-\frac{1}{2}}^{-}\right)_i^{t_i+w_i^*} \Phi^{l_i}_c(z), \quad w_i^*=l_i-2 t_i-w_i, \\
\left(U G_{-\frac{1}{2}}^{-}\right)_i^{t_i+w_i^*-l_i-1} U_i\left(U G_{-\frac{1}{2}}^{-}\right)_i^{l_i} \Phi^{l_i}_c(z), \quad w_i^*=k_i+2+l_i-2 t_i-w_i \simeq l_i-2 t_i -w_i.
\end{array}\right.
\end{aligned}
\label{eq:mirrorfield}
\end{equation}
The only difference is that this field appears as twisted by the element of another group, denoted $\check{G}$. This group is defined as follows:
\begin{equation}
    \label{eq:dualelement}
    \check{G}=\left\{\vec{w}^*=\vec{l}-2\vec{t}-\vec{w}=\vec{q}-\vec{w}\right\}.
\end{equation}
Recall that here, the vectors $\vec{q}$ label all possible charges of fields in composite model $M_{\vec{k}}$ that lie in the twisted sector of the element $\vec{w}$ of the orbifold $M_{\vec{k}}/G_{\text{adm}}$. Thus, these fields satisfy the mutual locality condition (\ref{eq:mutual}).

The set of vectors in (\ref{eq:dualelement}) indeed forms a group, as the vectors $\vec{w}^*$ are additive under the operator algebra of the orbifold model. This additivity stems from the fact that the $\vec{q}$ correspond to conserved $U(1)$ charges of the initial composite model $M_{\vec{k}}$ (i.e charges of two fields add under fusion), while the vectors $\vec{w}$ are additive by definition. It is also clear, that the elements $\vec{w}^{*}$ obey (\ref{eq:adm}) and hence, the group $\check{G}$ is admissible because the primary fields, which correspond to $(3,0)$ and $(0,3)$ differential forms on CY manifold are among the set of mutually local fields. Since this is an admissible group, we will refer to it as $\check{G}\equiv G_{\text{adm}}^*$.
Hence, (\ref{eq:mirrorfield})  can be interpreted as a field in the orbifold model that is built by the dual group: 

\begin{equation}
   \Psi^{\vec{l},\vec{\bar{l}}}_{\vec{t}+\vec{w}^*,\vec{\bar{t}}}(z,\bar{z})\in M_{\vec{k}}/G_{\text{adm}}^*.
\end{equation}
This dual orbifold model is isomorphic to $M_{\vec{k}}/G_{\text{adm}}$ by the construction. Moreover, as we will see shortly, this model is mirror to the initial orbifold with group $G_{\text{adm}}$. 

Thus, we have described the mirror spectral flow construction of primary fields in $NS$ sector. To build the fields in $R$ sector we simply apply the spectral flow to $NS$ primary fields:

\begin{equation}
\begin{aligned}
\prod_i U_i^{\frac{1}{2}} \bar{U}_i^{\frac{1}{2}} \tilde{\Psi}_{t_i+w^{*}_i,\bar{t}_i}^{l_i,\bar{l}_i}(z, \bar{z}). 
\end{aligned}
\end{equation}

Now let's return to the mutual locality equations (\ref{eq:mutual}). Taking into account the definition of the element of the dual group (\ref{eq:dualelement}), the equations (\ref{eq:mutual}) can be rewritten as

\begin{equation}
\label{eq:mutual*}
    \sum_{i=1}^r \frac{w_{1i}w_{2i}^*+w_{2i}w_{1i}^*}{k_i+2} \in \mathbb{Z}, \quad \vec{w}_{1,2}\in G_{adm}, \quad \vec{w}^{*}_{1,2}\in G^{*}_{adm}.
\end{equation}
The (\ref{eq:mutual*}) must hold for arbitrary pairs of twists $\vec{w}_1,\vec{w}_1^*$ and $\vec{w}_2,\vec{w}_2^*$. Consequently, they remain valid even when, for instance $\vec{w}_2^*=0$. Hence, (\ref{eq:mutual*}) is equivalent to
\begin{equation}
\label{eq:mirrorgroup}
\sum_{i=1}^r \frac{w_i w_i^*}{k_i+2} \in \mathbb{Z}, \quad \vec{w} \in G_{\text{adm}}, \quad \vec{w}^* \in G_{\text{adm}}^*.
\end{equation}

\subsection{Dual group}
If the composite model consists only of minimal models of type $\textbf{A}$, the equations (\ref{eq:mirrorgroup}) are nothing else but the definition of Berglund-Hubsch-Krawitz dual group for the Fermat type potential $W=X_{1}^{k_{1}+2}+...+X_{r}^{k_{r}+2}$ \cite{BH,Kra,ABBE}, so in this case, the group $G_{\text{adm}}^{*}$ is Berglund-Hubsch-Krawitz dual to $G_{\text{adm}}$.

In the case where the composite model consists of minimal models of arbitrary types, equations (\ref{eq:mirrorgroup}) can be considered as defining the generalization of the Berglund-Hubsch-Krawitz dual group. More precisely, according to our discussion above, we define the dual group as
\begin{equation}
\label{eq:dualgroup}
    G_{\text{adm}}^*=\left\{\vec{w}^*=\vec{q}-\vec{w} \ \bigg| \  \vec{w}\in G_{\text{adm}} ; \quad  \sum_{i=1}^r\frac{w_{i}w^{*}_{i}}{k_i+2}\in \mathbb{Z}; \quad\sum_{i=1}^r\frac{w_i^*}{k_i+2}\in \mathbb{Z}\right\} \subset G_{\text{tot}}.
\end{equation}

Having the admissible group $G_{\text{adm}}$ fixed and finding all possible elements of the dual group from (\ref{eq:mirrorgroup}),  we determine the full set of charges for mutually local fields for the orbifold model $M_{\vec{k}}/G_{\text{adm}}$ and $M_{\vec{k}}/G_{\text{adm}}^*$:
\begin{equation}
    \vec{q}=\vec{w}+\vec{w}^*, \quad \vec{w} \in G_{\text{adm}}, \quad \vec{w}^* \in G_{\text{adm}}^*.
    \label{eq:charge}
\end{equation}
In other words, the set of mutually local fields of the model $M_{\vec{k}}/G_{\text{adm}}$, which appear in the twisted sector $\vec{w}$ is given by the elements $\vec{w}^*\in G_{\text{adm}}^*$. From the other hand, the set of mutually local fields of the model $M_{\vec{k}}/G_{\text{adm}}^*$, which appear in the twisted sector $\vec{w}^*$ is given by the elements $\vec{w}\in G_{\text{adm}}$.

Notice that these equations and hence, the construction of local fields in the mutually dual orbifold models $M_{\vec{k}}/G_{\text{adm}}$ and $M_{\vec{k}}/G_{\text{adm}}^*$, are consistent with the invariants of types $\textbf{D}$ and $\textbf{E}$. 

That is, in the component $M_{k_i}$, the charges $q_i$ take an admissible set of values. Specifically, for invariants of type $\textbf{A}_{k_i+1}$, $\textbf{D}_{2j+1}$, or $\textbf{E}_{6}$ in $M_{k_i}$, the charges $q_i$, as well as the elements $w_i,w_i^*$ of the admissible groups take all possible values modulo $k_i+2$. 
For invariants of type $\textbf{D}_{2j+2}$, $\textbf{E}_7$, or $\textbf{E}_8$, the elements $w_i,w_i^*$ remain defined modulo $k_i+2$, but are constrained to even integers: $w_i,w_i^*=0 \ (\text{mod} \ 2)$ if we consider them as acting by (\ref{eq:Gtot}). Therefore the elements $w_i,w_i^*$  effectively belong to the group $\mathbb{Z}_{(k_i+2)/2}$. At the same time, the admissible charges are also restricted to even ones, as we discussed in subsection \ref{sec:group}. From the other hand, restricting the values of elements $w_i \in G_{\text{adm}}$ and $w_i^* \in G_{\text{adm}}^*$ to even values in $\mathbb{Z}_{k_i+2}$ is necessary to ensure consistency of the action of $G_{\text{adm}}$ and $G_{\text{adm}}^*$ by the spectral flow twists with the isomorphism
$M_{\vec{k}}/G_{\text{adm}}\leftrightarrow M_{\vec{k}}/G_{\text{adm}}^*$.

We have thereby demonstrated that both the definition of the dual group and the construction of the dual orbifold model (mirror model) naturally extend to composite models with an arbitrary ADE modular invariant. 

\subsection{Mirror Symmetry of mutually dual orbifolds}

Although it is more or less clear, we show in this subsection that the dual models $M_{\vec{k}}/G_{\text{adm}}$ and $M_{\vec{k}}/G_{\text{adm}}^*$ are mutually mirror indeed. It is given by showing that the isomorphism of models we have constructed above leads to isomorphism of $(c,c)$ ($(a,c)$) ring of model $M_{\vec{k}}/G_{\text{adm}}^*$ and the $(a,c)$ ($(c,c)$) ring of the model $M_{\vec{k}}/G_{\text{adm}}$.

$(c,c)$ - fields for any of the orbifold models can be found using the following analysis. According to the discussion above, we construct the $NS$ orbifold fields by applying the spectral flow operator to the fields of the composite model only in the holomorphic sector.
Hence, the anti-holomorphic factor of the $(c,c)$ field of the orbifold must be the chiral primary state $\bar{\Phi}^{\vec{\bar{l}}}_c$  with $\vec{q}=\vec{\bar{l}}$. So, the complete $(c,c)$ field can be represented as
\begin{equation}
 \Psi_{\vec{t}+\vec{w},0}^{\vec{l},\vec{\bar{l}}}(z, \bar{z})=V_{\vec{t}+\vec{w}}^{\vec{l}}(z) \bar{\Phi}_{c}^{\vec{\bar{l}}}(\bar{z}).
 \label{eq:chiral}
\end{equation}   
Its holomorphic factor may be off-diagonal in $l_i$, but has to be chiral primary state also. It is not difficult to see from (\ref{eq:twistfield}) - (\ref{eq:rightstate}), that this happens if
\begin{equation}
t_i+w_i=0 \bmod k_i+2, \quad \text { or } \ t_i+w_i=l_i+1 \bmod k_i+2,
\label{eq:cchiral}
\end{equation}
where $t_{i}$ are restricted by the condition that before the twisting the holomorphic charge coincides with the anti-holomorphic charge: $q_{i}=l_{i}-2t_{i}=\bar{l}_{i}$ (since the composite models under consideration are subject to this condition).

By the similar arguments we find that complete $(a,c)$ field of the orbifold must have the form (\ref{eq:chiral}), with the same restrictions for $t_{i}$, but this time its holomorphic
factor has to be the anti-chiral field. This happens if
\begin{equation}
t_i+w_i=k_i+1 \bmod k_i+2, \quad \text { or } \  t_i+w_i=l_i \bmod k_i+2 .
\label{eq:acchiral}
\end{equation}

It is important to note here that, unlike the \textbf{A} series models, the states in the holomorphic sector of the composite model can already be twisted, with an untwisted state (i.e. chiral primary) in the anti-holomorphic one. That is why we have parameters $t_i$ in the formulas above.

Taking into account the spectral flow construction (\ref{eq:twistfield}) and its mirror version (\ref{eq:mirrorfield}), as well as the relation (\ref{eq:dualelement}) it is now easy to see that $(c,c)$-field $V^{\vec{l}}_{\vec{t}+\vec{w}}\bar{V}^{{\vec{\bar{l}}}}_{\vec{0}} $ in the original model arises as $(a,c)$-field $\tilde{V}^{\vec{l}}_{\vec{t}+\vec{w}^*}\bar{V}^{{\vec{\bar{l}}}}_{\vec{0}} $ in the dual model provided the following correspondence for the twisted vectors:
\begin{equation}
\label{eq:dualtwist1}
    \begin{aligned}
       & t_{i}+w_i=l_i+1 \quad \leftrightarrow \quad t_i+w_i^*=q_i+2t_i-l_i-1=-1\simeq k_i+1 \\
       & t_{i}+w_i=k_i+2 \quad \leftrightarrow \quad t_i+w_i^*=q_i+2t_i=l_i. 
    \end{aligned}
\end{equation}

Analogously we find that $(a,c)$ field in the original model corresponds to the $(c,c)$ field in the dual one if the twisted vectors relate as:
\begin{equation}
\label{eq:dualtwist2}
    \begin{aligned}
       & t_{i}+w_i=l_i \quad \leftrightarrow \quad t_i+w_i^*=q_i+2t_i-l_i=0 \\
       & t_{i}+w_i=k_i+1 \quad \leftrightarrow \quad t_i+w_i^*=q_i+2t_i-(k_i+1)\simeq l_i+1. 
    \end{aligned}
\end{equation}
In (\ref{eq:dualtwist1}), (\ref{eq:dualtwist2}) we have used that vectors $w_i$ are defined modulo $k_i+2$. 

Besides, these relations have been obtained under the assumption $|q_i|\leq l_i$.
If the charge exceeds $|q_i|>l_i$ we must identify the fields as in (\ref{eq:identification}). The charge becomes $\tilde{q}_i=q_i+(k_i+2)$, but this does not affect the condition on the dual twist, since $w_i^{*}=q_i-w_i\simeq q_i+(k_i+2)-w_i=\tilde{q}_i-w_i$. Then the conditions (\ref{eq:dualtwist1}),(\ref{eq:dualtwist2}) are rewritten with a replacement $\tilde{l}_i=k-l_i$, and $\tilde{q}_i=\tilde{l}_i-2\tilde{t}_i$. 

It finishes the proof of the statement.

\section{Examples of orbifold models with a non-diagonal modular invariants}
\label{sec:5}

Here we illustrate the construction of fields for several examples of orbifolds of the composite model with an off-diagonal modular invariant. More precisely, we consider two mirror pairs of orbifolds of the composite model $(1_A)^1(16_E)^3$, which is the product of one model with \textbf{A} modular invariant at $k=1$ and three $k=16$ models with $\textbf{E}_7$  modular invariant (\ref{eq:ADE}). 

The first mirror pair of orbifolds is given by the pair of mutually dual groups
\begin{equation}
    G=\langle (2,2,2,2)\rangle, \ G^*=\langle(2,2,2,2),(1,0,0,12),(1,2,4,6),(0,0,4,14)\rangle,    
\label{eq:pair1}
\end{equation}
where the vector $(2,2,2,2)$ is the generator of the first group, while the vectors $(2,2,2,2)$, $(1,0,0,12)$, $(1,2,4,6)$, $(0,0,4,14)$ are the generators of the second group. For this pair of models we write out the $(a,c)$-fields which are realized by the formulas (\ref{eq:twistfield})-(\ref{eq:leftstate1}). 

For the orbifold with group $G$ we relate these fields with De Rham cohomology of the Calabi-Yau manifold $X$ 
, which is defined as an intersection of two hypersurfaces in $\mathbb{P}^2\times\mathbb{P}^3$ \cite{Schimmrigk}:
\begin{equation}
\label{eq:CY3gen}
    \begin{aligned}
         W_1(x)&=x_0^3+x_1^3+x_2^3+x_3^3=0 \\
        W_2(x,y)&=x_1y_1^3+x_2y_2^3+x_3y_3^3=0,
    \end{aligned}
\end{equation}
where $x_i\in \mathbb{P}^3$, and $y_i\in \mathbb{P}^2$. The Hodge numbers are $h^{2,1}(X)=h^{1,2}(X)=35, \quad h^{2,2}(X)=h^{1,1}(X)=8$, $h^{0,0}(X)=h^{0,3}(X)=h^{3,0}(X)=h^{3,3}(X)=1$. This orbifold model has been considered by D.Gepner \cite{Gepner:1987hi} as an initial step in his construction of three-generation model of superstring compactification. 

The $(a,c)$-fields of the mirror orbifold obviously coincide with De Rham cohomology of the mirror Calabi-Yau manifold, which is nothing but the orbifold of $X$ by "maximal admissible" group $G^*$ and at the same time, they are $(c,c)$-fields in the initial orbifold model, as can be seen directly from our construction.

The results are collected in tables \ref{tabular:2}, \ref{tabular:3}.
We use everywhere in tables the notation 
\begin{equation}
\Phi^{(l_1,l_2,l_3,l_4)}_{(q_1,q_2,q_3,q_4)}\bar{\Phi}^{(\bar{l}_1,\bar{l}_2,\bar{l}_3,\bar{l}_4)}_{(\bar{q}_1,\bar{q}_2,\bar{q}_3,\bar{q}_4)}:=\prod_i\Phi^{l_i}_{q_i}\bar{\Phi}^{\bar{l}_i}_{\bar{q}_i}.
\end{equation}

The structure of the tables is as follows. 
The first two columns of the tables contain the number of $(a,c)$ fields with charge $(Q_L,Q_R)$ and corresponding De-Rham cohomologies of CY. The untwisted fields from the composite model, which are located in the penultimate column, have the form $\Phi^{\vec{l}}_{\vec{q}}\bar{\Phi}^{\vec{\bar{l}}}_{c}$. The set of charges $\vec{q}$ of a holomorphic state coincides with the set of charges of an anti-holomorphic chiral state, that is $q_i=\bar{l}_i$. In the case when $|q_i|>l_i$ the corresponding field should be identified according to the (\ref{eq:identification}). In the "Field"
column are located the $(a,c)$ fields in the orbifold model, twisting by the element $\vec{w}$ from the last column.



The second mirror pair of orbifolds is given by the pair of mutually dual groups
\begin{equation}
    \tilde{G}=\langle (2,2,2,2),(0,6,12,0)\rangle, \ \tilde{G}^*=\langle(2,2,2,2),(0,4,4,10),(0,4,10,4)\rangle,  
\label{eq:pair2}
\end{equation}
where again, we defined the groups by their generators.

We write out again the $(a,c)$-fields of the first orbifold and relate them with De Rham cohomology groups of the Calabi-Yau manifold $Y$, which is defined as an orbifold of the intersection of hypersurfaces (\ref{eq:CY3gen}) by the group $\tilde{G}$ in $\mathbb{P}^2\times\mathbb{P}^3/\tilde{G}$:
 The Hodge numbers are $h^{2,1}(Y)=h^{1,2}(Y)=23, \quad h^{2,2}(Y)=h^{1,1}(Y)=14$, $h^{0,0}(Y)=h^{0,3}(Y)=h^{3,0}(Y)=h^{3,3}(Y)=1$.

The $(a,c)$-fields of the mirror orbifold correspond to De Rham cohomology of the mirror Calabi-Yau manifold which is defined as the orbifold of the original one (\ref{eq:CY3gen}) by dual group $\tilde{G}^*$. From the other hand, as can be seen directly from our construction, these fields consist of the $(c,c)$ fields of $\tilde{G}$ orbifold.

The results are given in tables \ref{tabular:4},\ref{tabular:5}.

As we can see, every orbifold model we have considered here, contains holomorphic and anti-holomorphic currents of charge 3, corresponding to the nowhere-vanishing (3,0) and (0,3) forms on the Calabi-Yau manifold. As can be directly checked, the mutual locality of these currents with the other fields of the model enforces the admissible group constraint (\ref{eq:CYcondition}) introduced in Section \ref{sec:3}. It can be shown that this is a general result: for orbifold model $M_{\vec{k}}/G_{\text{adm}}$ composed of minimal models $M_{k_i}$ with modular invariants from ADE list, the mutual locality of the fields with charges $(\pm3,0)$ and $(0,\pm3)$ with all other fields in the orbifold model is equivalent to condition \eqref{eq:CYcondition}.

Finally, we briefly discuss Gepner’s three-generation model.
To obtain the model with three generations, we need to make the orbifold over the group $\tilde{G}$ and factorise by the cyclic permutations $S\simeq \mathbb{Z}_3$. The group $S$ acts on three $\mathbf{E}_7$ models, permuting them cyclically. The resulting model $(1_A)^1(16_E)^3/(\tilde{G}\times S)$ has 9 generations (chiral-chiral fields) and 6 anti-generations (anti-chiral - chiral fields). Thus we get the explicit construction of fields in the compact sector of Gepner's three generations model of superstring compactification. 

\section{Conclusion}

We presented the explicit construction of fields in orbifolds of products of $N=(2,2)$ minimal models of ADE types, using spectral flow twisting and conformal bootstrap axioms. The spaces of fields of the initial orbifold model $M_{\vec{k}}/G_{\text{adm}}$ and their mirror $M_{\vec{k}}/G_{\text{adm}}^*$ are shown to be isomorphic, with the groups $G_{\text{adm}}$ and $G_{\text{adm}}^*$ related by Berglund-Hubsch-Krawitz-like condition. We shown that the spaces of fields of the initial orbifold $M_{\vec{k}}/G_{\text{adm}}$ and its mirror $M_{\vec{k}}/G_{\text{adm}}^*$ appear in the construction simultaneously. In other words, the mirror symmetry isomorphism of fields is built into the conformal bootstrap construction, as was suggested initially in \cite{Lerche:1989uy}. Our results can also be considered as a conformal bootstrap version of the modular bootstrap construction of mirror orbifolds from \cite{GrPl}. The geometric interpretation of the orbifolds of ADE models that we consider was studied in \cite{Fuchs:1989pt, Lynker}, where it was shown that such orbifolds are related to Calabi-Yau manifolds defined as a complete intersection in projective spaces. In this paper, we thereby have given an explicit construction of fields in $\sigma$-models on the Calabi-Yau manifolds studied there.

The approach has been illustrated through the example of mirror pairs of orbifolds of the $(1_A)(16_E)^3$ model. 

A natural extension of this work would be to generalize our explicit construction of fields to the orbifolds of products of minimal models with the broader class of modular invariants found in \cite{Gannon:1996hp}.

Our explicit construction of fields of the orbifolds can certainly be extended to build the physical states in the models of Type $II$ and Heterotic superstring compactifications. The isomorphism mapping between the mirror orbifold models enables a proof of mirror symmetry of states between the $IIA$ and $IIB$ models compactified on the orbifolds of ADE-type composite models, similar to what is done in \cite{Parkhomenko:2024mxq}. 

Furthermore, we hope that the explicit field construction developed here will prove useful for computing correlation functions in the approach advocated in \cite{FST}.

\section*{Acknowledgments}

We are grateful to prof. A. Belavin, and to A. Litvinov for useful discussions. 
The work of B. Eremin was carried out within the state assignment of Ministry of Science and Higher Education of the Russian Federation for IITP RAS.  The work of S. Parkhomenko was
carried out at Landau Institute for Theoretical Physics in the framework of the state assignment
FFWR-2024-0012.

\newpage

\appendix
\section{Tables}

\begin{table}[H]

\caption{Spectrum of $E_7$ Minimal Model. $NS$-sector}
\setlength{\tabcolsep}{10pt} 
\renewcommand{\arraystretch}{1.5} 
\label{tabular:E7}
\begin{center}
\begin{tabular}{ |c|c|c|} 
 \hline
 Term in $Z_{E_7}$ & Field: $\Phi_{q}^l\bar{\Phi}_{\bar{q}}^{\bar{l}}$ & Range of $q$ (only even values) \\
\hline
 \hline 

$\left|\chi_0+\chi_{16}\right|^2$ & $\Phi_{0}^0\bar{\Phi}_{0}^{16}$, $\Phi_{0}^{16}\bar{\Phi}_{0}^0$ & \\ 
& $\Phi_{0}^0\bar{\Phi}_{0}^0$ & \\ 
& $\Phi_{q}^{16}\bar{\Phi}_{q}^{16}$ & $q=-16,\dots,16$ \\
& $\Phi^{16}_{q-18}\bar{\Phi}_{q}^{16}$, $\Phi_{q}^{16}\bar{\Phi}_{q-18}^{16}$  & $q=2,\dots,16$ \\

\hline
 \hline 
$\left|\chi_4+\chi_{12}\right|^2$ & $\Phi_{q}^4\bar{\Phi}_{q}^{12}$, $\Phi_{q}^{12}\bar{\Phi}_{q}^4$& $q=-4,\dots,4$ \\ 
& $\Phi_{q}^4\bar{\Phi}_{q}^4$ & $q=-4,\dots,4$ \\
& $\Phi_{q}^{12}\bar{\Phi}_{q}^{12}$ & $q=-12,\dots,12$ \\
& $\Phi_{q-18}^{12}\bar{\Phi}_{q}^{12}$, $\Phi_{q}^{12}\bar{\Phi}_{q-18}^{12}$ & $q=6,\dots,12$ \\
\hline
 \hline 
$\left|\chi_6+\chi_{10}\right|^2$ & $\Phi_{q}^6\bar{\Phi}_{q}^{10}$, $\Phi_{q}^{10}\bar{\Phi}_{q}^6$ & $q=-6,\dots,6$ \\ 
& $\Phi_{q}^6\bar{\Phi}_{q}^6$ & $q=-6,\dots,6$ \\ 
& $\Phi_{q}^{10}\bar{\Phi}_{q}^{10}$ & $q=-10,\dots,10$ \\ 
& $\Phi_{q-18}^{10}\bar{\Phi}_{q}^{10}$, $\Phi_{q}^{10}\bar{\Phi}_{q-18}^{10}$ & $q=8,10$ \\
\hline
 \hline 
 $\left|\chi_8\right|^2$ &  $\Phi_{q}^8\bar{\Phi}_{q}^8$ & $q=-8,\dots,8$\\ 
 \hline
 \hline 
$\left(\chi_2+\chi_{14}\right)  \bar{\chi}_8+\chi_8\left(\bar{\chi}_2+\bar{\chi}_{14}\right)$ & $\Phi_{q}^2\bar{\Phi}_{q}^8$, $\Phi_{q}^8\bar{\Phi}_{q}^2$  & $q=-2,\dots,2$ \\
& $\Phi_{q}^{14}\bar{\Phi}_{q}^8$, $\Phi_{q}^8\bar{\Phi}_{q}^{14}$  & $q=-8,\dots,8$ \\
& $\Phi_{q-18}^{14}\bar{\Phi}_{q}^8$, $\Phi_{q}^8\bar{\Phi}_{q-18}^{14}$  & $q=4,\dots,8$ \\
& $\Phi_{q+18}^{14}\bar{\Phi}_{q}^{8}$, $\Phi_{q}^8\bar{\Phi}_{q+18}^{14}$  & $q=-4,\dots,-8$ \\
\hline
 \hline 

\end{tabular}
\end{center}
\end{table}

\begin{table}[H]
\caption{$(a,c)$ fields in $(1_A)^1(16_E)^3$ with  $G=\langle(2,2,2,2)\rangle$.}
\label{tabular:2}
\begin{center}
\begin{tabular}{ |c|c|c|c|c|c|c|c|c|} 
 \hline
 Nmb. & $H^{p,q}$ & $(Q_L,Q_R)$ & \# & Field: $\Phi^{\vec{l}}_{a}\bar{\Phi}^{\vec{\bar{l}}}_{c}$ & Obtained from: $\Phi^{\vec{l}}_{\vec{q}}\bar{\Phi}^{\vec{\bar{l}}}_{c}$ & $\vec{w}\in G$ \\
 \hline
 \hline

 1& $H^{0,0}$ & (0,0) & 1 &  $I\times\bar{I}$ & Untwisted &  \\

 \hline 

  8 & $H^{1,1}$ & (-1,1) & 1 &  $\Phi^{(0,6,6,6)}_a\bar{\Phi}^{(0,6,6,6)}_c$ & $\Phi^{(0,6,6,6)}_c\bar{\Phi}^{(0,6,6,6)}_c$ & $(0,6,6,6)$ \\

 &  & (-1,1) & 1 &  $\Phi^{(1,4,4,4)}_a\bar{\Phi}^{(1,4,4,4)}_c$ & $\Phi^{(1,4,4,4)}_c\bar{\Phi}^{(1,4,4,4)}_c$ & $(1,4,4,4)$ \\

 &  & (-1,1) & 3 &  $\Phi^{(1,8,2,2)}_a\bar{\Phi}^{(0,2,8,8)}_c$ & $\Phi^{(0,8,14,14)}_{(0,2,8,8)}\bar{\Phi}^{(0,2,8,8)}_c$ & $(2,14,14,14)$ \\

 &  & (-1,1) & 3 &  $\Phi^{(0,8,8,2)}_a\bar{\Phi}^{(1,2,2,8)}_c$ & $\Phi^{(1,8,8,14)}_{(1,2,2,8)}\bar{\Phi}^{(1,2,2,8)}_c$ & $(2,14,14,14)$ \\

\hline

1&  $H^{0,3}$ & (0,3) &  1 & $I\times\bar{\Phi}^{(1,16,16,16)}_c$ & $\Phi^{(1,0,0,0)}_{(1,16,16,16)}\bar{\Phi}^{(1,16,16,16)}_c$ & $(2,8,8,8)$ \\

 \hline

    35& $H^{1,2}$ & (-1,2) & 1 &  $\Phi^{(0,6,6,6)}_a\bar{\Phi}^{(1,10,10,10)}_c$ & $\Phi^{(1,6,6,6)}_{(1,10,10,10)}\bar{\Phi}^{(1,10,10,10)}_c$ & $(2,8,8,8)$ \\

   & &(-1,2) & 1 &  $\Phi^{(1,4,4,4)}_a\bar{\Phi}^{(0,12,12,12)}_c$ & $\Phi^{(0,4,4,4)}_{(0,12,12,12)}\bar{\Phi}^{(0,12,12,12)}_c$ & $(2,8,8,8)$ \\

   & &(-1,2) & 3 &  $\Phi^{(1,12,0,0)}_a\bar{\Phi}^{(0,4,16,16)}_c$ & $\Phi^{(0,12,0,0)}_{(0,4,16,16)}\bar{\Phi}^{(0,4,16,16)}_c$ & $(2,8,8,8)$ \\

   & &(-1,2) & 3 &  $\Phi^{(0,10,4,4)}_a\bar{\Phi}^{(1,6,12,12)}_c$ & $\Phi^{(1,10,4,4)}_{(1,6,12,12)}\bar{\Phi}^{(1,6,12,12)}_c$ & $(2,8,8,8)$ \\

   & &(-1,2) & 3 &  $\Phi^{(1,0,6,6)}_a\bar{\Phi}^{(0,16,10,10)}_c$ & $\Phi^{(0,0,6,6)}_{(0,16,10,10)}\bar{\Phi}^{(0,16,10,10)}_c$ & $(2,8,8,8)$ \\
    
   & &(-1,2) & 6 &  $\Phi^{(1,8,4,0)}_a\bar{\Phi}^{(0,8,12,16)}_c$ & $\Phi^{(0,8,4,0)}_{(0,8,12,16)}\bar{\Phi}^{(0,8,12,16)}_c$ & $(2,8,8,8)$ \\

   & &(-1,2) & 6 &  $\Phi^{(0,12,6,0)}_a\bar{\Phi}^{(1,4,10,16)}_c$ & $\Phi^{(1,12,6,0)}_{(1,4,10,16)}\bar{\Phi}^{(1,4,10,16)}_c$ & $(2,8,8,8)$ \\

   & &(-1,2) & 6 &  $\Phi^{(0,10,8,0)}_a\bar{\Phi}^{(1,6,8,16)}_c$ & $\Phi^{(1,10,8,0)}_{(1,6,8,16)}\bar{\Phi}^{(1,6,8,16)}_c$ & $(2,8,8,8)$ \\

   & & (-1,2) & 6 &  $\Phi^{(0,8,6,4)}_a\bar{\Phi}^{(1,8,10,12)}_c$ & $\Phi^{(1,8,6,4)}_{(1,8,10,12)}\bar{\Phi}^{(1,8,10,12)}_c$ & $(2,8,8,8)$ \\

  \hline

 35& $H^{2,1}$ & (-2,1) & 1 &  $\Phi^{(1,10,10,10)}_a\bar{\Phi}^{(0,6,6,6)}_c$ & $\Phi^{(0,10,10,10)}_{(0,6,6,6)}\bar{\Phi}^{(0,6,6,6)}_c$ & $(2,8,8,8)$ \\

  & &(-2,1) & 1 &  $\Phi^{(0,12,12,12)}_a\bar{\Phi}^{(1,4,4,4)}_c$ & $\Phi^{(1,12,12,12)}_{(1,4,4,4)}\bar{\Phi}^{(1,4,4,4)}_c$ & $(2,8,8,8)$ \\
 
  & &(-2,1) & 3 &  $\Phi^{(0,4,16,16)}_a\bar{\Phi}^{(1,12,0,0)}_c$ & $\Phi^{(1,4,16,16)}_{(1,12,0,0)}\bar{\Phi}^{(1,12,0,0)}_c$ & $(2,8,8,8)$ \\

  & &(-2,1) & 3 &  $\Phi^{(1,6,12,12)}_a\bar{\Phi}^{(0,10,4,4)}_c$ & $\Phi^{(0,6,12,12)}_{(0,10,4,4)}\bar{\Phi}^{(0,10,4,4)}_c$ & $(2,8,8,8)$ \\

  & &(-2,1) & 3 &  $\Phi^{(0,16,10,10)}_a\bar{\Phi}^{(1,0,6,6)}_c$ & $\Phi^{(1,16,10,10)}_{(1,0,6,6)}\bar{\Phi}^{(1,0,6,6)}_c$ & $(2,8,8,8)$ \\

  & &(-2,1) & 6 &  $\Phi^{(0,8,12,16)}_a\bar{\Phi}^{(1,8,4,0)}_c$ & $\Phi^{(1,8,12,16)}_{(1,8,4,0)}\bar{\Phi}^{(1,8,4,0)}_c$ & $(2,8,8,8)$ \\

  & &(-2,1) & 6 &  $\Phi^{(1,4,10,16)}_a\bar{\Phi}^{(0,12,6,0)}_c$ & $\Phi^{(0,4,10,16)}_{(0,12,6,0)}\bar{\Phi}^{(0,12,6,0)}_c$ & $(2,8,8,8)$ \\

  & &(-2,1) & 6 &  $\Phi^{(1,6,8,16)}_a\bar{\Phi}^{(0,10,8,0)}_c$ & $\Phi^{(0,6,8,16)}_{(0,10,8,0)}\bar{\Phi}^{(0,10,8,0)}_c$ & $(2,8,8,8)$ \\

  & &(-2,1) & 6 &  $\Phi^{(1,8,10,12)}_a\bar{\Phi}^{(0,8,6,4)}_c$ & $\Phi^{(0,8,10,12)}_{(0,8,6,4)}\bar{\Phi}^{(0,8,6,4)}_c$ & $(2,8,8,8)$ \\

\hline

1& $H^{3,0}$ & (-3,0) & 1 &  $\Phi^{(1,16,16,16)}_a\times\bar{I}$ & $\Phi^{(0,16,16,16)}_{(0,0,0,0)}\times\bar{I}$ & $(2,8,8,8)$ \\

\hline

   8& $H^{2,2}$ & (-2,2) & 1 &  $\Phi^{(0,12,12,12)}_a\bar{\Phi}^{(0,12,12,12)}_c$ & $\Phi^{(0,12,12,12)}_c\bar{\Phi}^{(0,12,12,12)}_c$ & $(0,12,12,12)$ \\
  
  & & (-2,2) & 1 &  $\Phi^{(1,10,10,10)}_a\bar{\Phi}^{(1,10,10,10)}_c$ & $\Phi^{(1,10,10,10)}_c\bar{\Phi}^{(1,10,10,10)}_c$ & $(1,10,10,10)$ \\

 & & (-2,2) & 3 &  $\Phi^{(0,14,14,8)}_a\bar{\Phi}^{(1,8,8,14)}_c$ & $\Phi^{(1,2,2,8)}_{(1,8,8,14)}\bar{\Phi}^{(1,8,8,14)}_c$ & $(2,2,2,2)$ \\

  & & (-2,2) & 3  &  $\Phi^{(1,14,8,8)}_a\bar{\Phi}^{(0,8,14,14)}_c$ & $\Phi^{(0,2,8,8)}_{(0,8,14,14)}\bar{\Phi}^{(0,8,14,14)}_c$ & $(2,2,2,2)$ \\

\hline

1 & $H^{3,3}$ & (-3,3) &  1 &$\Phi^{(1,16,16,16)}_a\bar{\Phi}^{(1,16,16,16)}_c$ & $\Phi^{(1,16,16,16)}_c\bar{\Phi}^{(1,16,16,16)}_c$ & $(1,16,16,16)$ \\

 \hline

\end{tabular}
\end{center}
\end{table}


\begin{table}[H]
\caption{$(a,c)$ fields in the mirror of $(1_A)^1(16_E)^3$, the group is $G^*$.}
\label{tabular:3}
\begin{center}
\begin{tabular}{ |c|c|c|c|c|c|c|} 
 \hline
 Nmb. & $H^{p,q}$ & $(Q_L,Q_R)$ & \# & Field: $\Phi^{\vec{l}}_{a}\bar{\Phi}^{\vec{\bar{l}}}_{c}$ & Obtained from: $\Phi^{\vec{l}}_{\vec{q}}\bar{\Phi}^{\vec{\bar{l}}}_{c}$ & $\vec{w}^*\in G^*$   \\
 \hline
 \hline

 1& $H^{0,0}$ & (0,0) & 1 &  $I\times\bar{I}$ & Untwisted &  \\

 \hline 
 
35& $H^{1,1}$ & (-1,1) & 1 &  $\Phi_{a}^{(0,6,6,6)}\bar{\Phi}_{c}^{(0,6,6,6)}$ & $\Phi_{c}^{(0,6,6,6)}\bar{\Phi}_{c}^{(0,6,6,6)}$ &(0,6,6,6)  \\
& & (-1,1) & 1 &  $\Phi_{a}^{(1,4,4,4)}\bar{\Phi}_{c}^{(1,4,4,4)}$ & $\Phi_{c}^{(1,4,4,4)}\bar{\Phi}_{c}^{(1,4,4,4)}$ &(1,4,4,4)  \\
& &(-1,1) & 3 &  $\Phi_{a}^{(0,4,4,10)}\bar{\Phi}_{c}^{(0,4,4,10)}$ & $\Phi_{c}^{(0,4,4,10)}\bar{\Phi}_{c}^{(0,4,4,10)}$ &(0,4,4,10) \\
& &(-1,1) & 3 &  $\Phi_{a}^{(1,0,0,12)}\bar{\Phi}_{c}^{(1,0,0,12)}$ & $\Phi_{c}^{(1,0,0,12)}\bar{\Phi}_{c}^{(1,0,0,12)}$ &(1,0,0,12) \\
& &(-1,1) & 3 &  $\Phi_{a}^{(1,0,6,6)}\bar{\Phi}_{c}^{(1,0,6,6)}$ & $\Phi_{c}^{(1,0,6,6)}\bar{\Phi}_{c}^{(1,0,6,6)}$ &(1,0,6,6) \\

& &(-1,1) & 6 &  $\Phi_{a}^{(0,0,6,12)}\bar{\Phi}_{c}^{(0,0,6,12)}$ & $\Phi_{c}^{(0,0,6,12)}\bar{\Phi}_{c}^{(0,0,6,12)}$ &(0,0,6,12)   \\
& &(-1,1) & 6 &  $\Phi_{a}^{(0,0,8,10)}\bar{\Phi}_{c}^{(0,0,8,10)}$ & $\Phi_{c}^{(0,0,8,10)}\bar{\Phi}_{c}^{(0,0,8,10)}$ &(0,0,8,10)   \\
& &(-1,1) & 6 &  $\Phi_{a}^{(0,4,6,8)}\bar{\Phi}_{c}^{(0,4,6,8)}$ & $\Phi_{c}^{(0,4,6,8)}\bar{\Phi}_{c}^{(0,4,6,8)}$ &(0,4,6,8)  \\

& &(-1,1) & 6 &  $\Phi_{a}^{(1,0,4,8)}\bar{\Phi}_{c}^{(1,0,4,8)}$ & $\Phi_{c}^{(1,0,4,8)}\bar{\Phi}_{c}^{(1,0,4,8)}$ &(1,0,4,8) \\

 \hline
1&  $H^{0,3}$ & (0,3) &  1 & $I\times\bar{\Phi}^{(1,16,16,16)}_c$ & $\Phi^{(1,0,0,0)}_{(1,16,16,16)}\bar{\Phi}^{(1,16,16,16)}_c$ & $(2,8,8,8)$ \\

 \hline

  8& $H^{1,2}$ & (-1,2) & 1 &  $\Phi^{(0,6,6,6)}_a\bar{\Phi}^{(1,10,10,10)}_c$ & $\Phi^{(1,6,6,6)}_{(1,10,10,10)}\bar{\Phi}^{(1,10,10,10)}_c$ & $(2,8,8,8)$ \\

  & & (-1,2) & 1 &  $\Phi^{(1,4,4,4)}_a\bar{\Phi}^{(0,12,12,12)}_c$ & $\Phi^{(0,4,4,4)}_{(0,12,12,12)}\bar{\Phi}^{(0,12,12,12)}_c$ & $(2,8,8,8)$ \\
 
  & & (-1,2) & 3 &  $\Phi^{(1,2,2,8)}_{a}\bar{\Phi}^{(1,8,8,14)}_c$ & $\Phi^{(1,14,14,8)}_{(1,8,8,14)}\bar{\Phi}^{(1,8,8,14)}_c$ & $(1,14,14,2)$ \\

  &  & (-1,2) & 3  &  $\Phi^{(0,2,8,8)}_a\bar{\Phi}^{(0,8,14,14)}_c$ & $\Phi^{(0,14,8,8)}_{(0,8,14,14)}\bar{\Phi}^{(0,8,14,14)}_c$ & $(0,12,12,12)$ \\

 \hline

   8 & $H^{2,1}$  & (-2,1) & 1 &  $\Phi^{(1,10,10,10)}_a\bar{\Phi}^{(0,6,6,6))}_c$ & $\Phi^{(0,10,10,10)}_{(0,6,6,6)}\bar{\Phi}^{(0,6,6,6)}_c$ & $(2,8,8,8)$ \\

  & & (-2,1) & 1 &  $\Phi^{(0,12,12,12)}_a\bar{\Phi}^{(1,4,4,4)}_c$ & $\Phi^{(1,12,12,12)}_{(1,4,4,4)}\bar{\Phi}^{(1,4,4,4)}_c$ & $(2,8,8,8)$ \\
 
  & & (-2,1) & 3 &  $\Phi^{(1,8,8,14)}_a\bar{\Phi}^{(1,2,2,8)}_c$ & $\Phi^{(1,8,8,2)}_{(1,2,2,8)}\bar{\Phi}^{(1,2,2,8)}_c$ & $(1,14,14,2)$ \\

  & & (-2,1) & 3  &  $\Phi^{(0,8,14,14)}_a\bar{\Phi}^{(0,2,8,8)}_c$ & $\Phi^{(0,8,2,2)}_{(0,2,8,8)}\bar{\Phi}^{(0,2,8,8)}_c$ & $(0,14,2,2)$ \\
 \hline

1& $H^{3,0}$ & (-3,0) & 1 &  $\Phi^{(1,16,16,16)}_a\times\bar{I}$ & $\Phi^{(0,16,16,16)}_{(0,0,0,0)}\times\bar{I}$ & $(2,8,8,8)$ \\

 \hline

   35& $H^{2,2}$ & (-2,2) & 1 &  $\Phi_{a}^{(0,12,12,12)}\bar{\Phi}_{c}^{(0,12,12,12)}$ & $\Phi_{c}^{(0,12,12,12)}\bar{\Phi}_{c}^{(0,12,12,12)}$ & (0,12,12,12) \\
& & (-2,2) & 1 &  $\Phi_{a}^{(1,10,10,10)}\bar{\Phi}_{c}^{(1,10,10,10)}$  & $\Phi_{c}^{(1,10,10,10)}\bar{\Phi}_{c}^{(1,10,10,10)}$ & (1,10,10,10) \\
& & (-2,2) & 3 &  $\Phi_{a}^{(0,4,16,16)}\bar{\Phi}_{c}^{(0,4,16,16)}$ & $\Phi_{c}^{(0,4,16,16)}\bar{\Phi}_{c}^{(0,4,16,16)}$ & (0,4,16,16) \\
& & (-2,2) & 3 &  $\Phi_{a}^{(0,10,10,16)}\bar{\Phi}_{c}^{(0,10,10,16)}$ & $\Phi_{c}^{(0,10,10,16)}\bar{\Phi}_{c}^{(0,10,10,16)}$&(0,10,10,16) \\
& & (-2,2) & 3 &  $\Phi_{a}^{(1,6,12,12)}\bar{\Phi}_{c}^{(1,6,12,12)}$ & $\Phi_{c}^{(1,6,12,12)}\bar{\Phi}_{c}^{(1,6,12,12)}$ &(1,6,12,12) \\

& & (-2,2) & 6 &  $\Phi_{a}^{(0,8,12,16)}\bar{\Phi}_{c}^{(0,8,12,16)}$ & $\Phi_{c}^{(0,8,12,16)}\bar{\Phi}_{c}^{(0,8,12,16)}$& (0,8,12,16)\\
& &(-2,2) & 6 &  $\Phi_{a}^{(1,4,10,16)}\bar{\Phi}_{c}^{(1,4,10,16)}$ & $\Phi_{c}^{(1,4,10,16)}\bar{\Phi}_{c}^{(1,4,10,16)}$ & (1,4,10,16) \\
& &(-2,2) & 6 &  $\Phi_{a}^{(1,6,8,16)}\bar{\Phi}_{c}^{(1,6,8,16)}$ & $\Phi_{c}^{(1,6,8,16)}\bar{\Phi}_{c}^{(1,6,8,16)}$ & (1,6,8,16)\\

& &(-2,2) & 6 &  $\Phi_{a}^{(1,8,10,12))}\bar{\Phi}_{c}^{(1,8,10,12)}$ & $\Phi_{c}^{(1,8,10,12))}\bar{\Phi}_{c}^{(1,8,10,12)}$ & (1,8,10,12) \\

\hline
 1 & $H^{3,3}$ & (-3,3) &  1 &$\Phi^{(1,16,16,16)}_a\bar{\Phi}^{(1,16,16,16)}_c$ & $\Phi^{(1,16,16,16)}_c\bar{\Phi}^{(1,16,16,16)}_c$ & $(1,16,16,16)$ \\
\hline 

\end{tabular}
\end{center}
\end{table}

\begin{table}[H]
\caption{$(a,c)$ fields of $(1_A)^1(16_E)^3$ with $\tilde{G}=\langle (2,2,2,2), (0,6,12,0)\rangle$. }
\label{tabular:4}
\begin{center}
\begin{tabular}{ |c|c|c|c|c|c|c|} 
 \hline
 Nmb.& $H^{p,q}$ & $(Q_L,Q_R)$ & \# & Field: $\Phi^{\vec{l}}_{a}\bar{\Phi}^{\vec{\bar{l}}}_{c}$ & Obtained from: $\Phi^{\vec{l}}_{\vec{q}}\bar{\Phi}^{\vec{\bar{l}}}_{c}$ & $\vec{w}\in \tilde{G}$   \\
 \hline
 \hline

    1& $H^{0,0}$ & (0,0) & 1 &  $I\times\bar{I}$ & Untwisted &  \\

     \hline 
     
 14 & $H^{1,1}$ & (-1,1) & 1 &  $\Phi_{a}^{(0,6,6,6)}\bar{\Phi}_{c}^{(0,6,6,6)}$ & $\Phi_{c}^{(0,6,6,6)}\bar{\Phi}_{c}^{(0,6,6,6)}$ & $(0,6,6,6)$ \\

 & &(-1,1) & 1 &  $\Phi_{a}^{(1,4,4,4)}\bar{\Phi}_{c}^{(1,4,4,4)}$ & $\Phi_{c}^{(1,4,4,4)}\bar{\Phi}_{c}^{(1,4,4,4)}$ & $(1,4,4,4)$ \\

 & &(-1,1) & 3 &  $\Phi_{a}^{(1,8,2,2)}\bar{\Phi}_{c}^{(0,2,8,8)}$ & $\Phi^{(0,8,14,14)}_{(0,2,8,8)}\bar{\Phi}_{c}^{(0,2,8,8)}$ & $(2,14,14,14)$ \\

 & &(-1,1) & 3 &  $\Phi_{a}^{(0,8,8,2)}\bar{\Phi}_{c}^{(1,2,2,8)}$ & $\Phi^{(1,8,8,14)}_{(1,2,2,8)}\bar{\Phi}_{c}^{(1,2,2,8)}$ & $(2,14,14,14)$ \\

 & &(-1,1) & 6 &  $\Phi_{a}^{(0,6,12,0)}\bar{\Phi}_{c}^{(0,6,12,0)}$ & $\Phi_{c}^{(0,6,12,0)}\bar{\Phi}_{c}^{(0,6,12,0)}$ &(0,6,12,0)   \\

     \hline
    1&  $H^{0,3}$ & (0,3) &  1 & $I\times\bar{\Phi}^{(1,16,16,16)}_c$ & $\Phi^{(1,0,0,0)}_{(1,16,16,16)}\bar{\Phi}^{(1,16,16,16)}_c$ & $(2,8,8,8)$ \\
    
\hline

   23 & $H^{1,2}$ & (-1,2) & 1 &  $\Phi^{(0,6,6,6)}_a\bar{\Phi}^{(1,10,10,10)}_c$ & $\Phi^{(1,6,6,6)}_{(1,10,10,10)}\bar{\Phi}^{(1,10,10,10)}_c$ & $(2,8,8,8)$ \\

   & &(-1,2) & 1 &  $\Phi^{(1,4,4,4)}_a\bar{\Phi}^{(0,12,12,12)}_c$ & $\Phi^{(0,4,4,4)}_{(0,12,12,12)}\bar{\Phi}^{(0,12,12,12)}_c$ & $(2,8,8,8)$ \\

   & &(-1,2) & 3 &  $\Phi^{(1,12,0,0)}_a\bar{\Phi}^{(0,4,16,16)}_c$ & $\Phi^{(0,12,0,0)}_{(0,4,16,16)}\bar{\Phi}^{(0,4,16,16)}_c$ & $(2,8,8,8)$ \\

   & &(-1,2) & 3 &  $\Phi^{(0,10,4,4)}_a\bar{\Phi}^{(1,6,12,12)}_c$ & $\Phi^{(1,10,4,4)}_{(1,6,12,12)}\bar{\Phi}^{(1,6,12,12)}_c$ & $(2,8,8,8)$ \\

   & &(-1,2) & 3 &  $\Phi^{(1,0,6,6)}_a\bar{\Phi}^{(0,16,10,10)}_c$ & $\Phi^{(0,0,6,6)}_{(0,16,10,10)}\bar{\Phi}^{(0,16,10,10)}_c$ & $(2,8,8,8)$ \\

   & &(-1,2) & 6 &  $\Phi^{(0,12,6,0)}_a\bar{\Phi}^{(1,4,10,16)}_c$ & $\Phi^{(1,12,6,0)}_{(1,4,10,16)}\bar{\Phi}^{(1,4,10,16)}_c$ & $(2,8,8,8)$ \\

   & &(-1,2) & 3 &  $\Phi^{(0,8,2,8)}_a\bar{\Phi}^{(1,8,8,14)}_c$ & $\Phi^{(1,8,14,8)}_{(1,8,8,14)}\bar{\Phi}^{(1,8,8,14)}_c$ & $(2,8,14,2)$ \\

   & &(-1,2) & 3 &  $\Phi^{(0,2,8,8)}_a\bar{\Phi}^{(1,8,8,14)}_c$ & $\Phi^{(1,14,8,8)}_{(1,8,8,14)}\bar{\Phi}^{(1,8,8,14)}_c$ & $(2,14,8,2)$ \\

      \hline
        23& $H^{2,1}$ & (-2,1) & 1 &  $\Phi^{(1,10,10,10)}_a\bar{\Phi}^{(0,6,6,6)}_c$ & $\Phi^{(0,10,10,10)}_{(0,6,6,6)}\bar{\Phi}^{(0,6,6,6)}_c$ & $(2,8,8,8)$ \\

  & &(-2,1) & 1 &  $\Phi^{(0,12,12,12)}_a\bar{\Phi}^{(1,4,4,4)}_c$ & $\Phi^{(1,12,12,12)}_{(1,4,4,4)}\bar{\Phi}^{(1,4,4,4)}_c$ & $(2,8,8,8)$ \\
 
  & &(-2,1) & 3 &  $\Phi^{(0,4,16,16)}_a\bar{\Phi}^{(1,12,0,0)}_c$ & $\Phi^{(1,4,16,16)}_{(1,12,0,0)}\bar{\Phi}^{(1,12,0,0)}_c$ & $(2,8,8,8)$ \\

  & &(-2,1) & 3 &  $\Phi^{(1,6,12,12)}_a\bar{\Phi}^{(0,10,4,4)}_c$ & $\Phi^{(0,6,12,12)}_{(0,10,4,4)}\bar{\Phi}^{(0,10,4,4)}_c$ & $(2,8,8,8)$ \\

  & &(-2,1) & 3 &  $\Phi^{(0,16,10,10)}_a\bar{\Phi}^{(1,0,6,6)}_c$ & $\Phi^{(1,16,10,10)}_{(1,0,6,6)}\bar{\Phi}^{(1,0,6,6)}_c$ & $(2,8,8,8)$ \\

  & &(-2,1) & 6 &  $\Phi^{(1,4,10,16)}_a\bar{\Phi}^{(0,12,6,0)}_c$ & $\Phi^{(0,4,10,16)}_{(0,12,6,0)}\bar{\Phi}^{(0,12,6,0)}_c$ & $(2,8,8,8)$ \\

  & &(-2,1) & 3 &  $\Phi^{(1,8,14,8)}_a\bar{\Phi}^{(0,8,8,2)}_c$ & $\Phi^{(0,8,2,8)}_{(0,8,8,2)}\bar{\Phi}^{(0,8,8,2)}_c$ & $(2,8,2,14)$ \\

  & &(-2,1) & 3 &  $\Phi^{(1,14,8,8)}_a\bar{\Phi}^{(0,8,8,2)}_c$ & $\Phi^{(0,2,8,8)}_{(0,8,8,2)}\bar{\Phi}^{(0,8,8,2)}_c$ & $(2,2,8,14)$ \\

 \hline

    1& $H^{3,0}$ & (-3,0) & 1 &  $\Phi^{(1,16,16,16)}_a\times\bar{I}$ & $\Phi^{(0,16,16,16)}_{(0,0,0,0)}\times\bar{I}$ & $(2,8,8,8)$ \\
    \hline

14 & $H^{2,2}$ & (-2,2) & 1 &  $\Phi_{a}^{(0,12,12,12)}\bar{\Phi}_{c}^{(0,12,12,12)}$ & $\Phi_{c}^{(0,12,12,12)}\bar{\Phi}_{c}^{(0,12,12,12)}$ & $(0,12,12,12)$ \\
&  & (-2,2) & 1 &  $\Phi_{a}^{(1,10,10,10)}\bar{\Phi}_{c}^{(1,10,10,10)}$ & $\Phi_{c}^{(1,10,10,10)}\bar{\Phi}_{c}^{(1,10,10,10)}$ & $(1,10,10,10)$ \\
&  & (-2,2) & 3 &  $\Phi_{a}^{(0,14,14,8)}\bar{\Phi}_{c}^{(1,8,8,14)}$ & $\Phi_{(1,8,8,14)}^{(1,2,2,8)}\bar{\Phi}_{c}^{(1,8,8,14)}$ & $(2,2,2,2)$ \\
&  & (-2,2) & 3 &  $\Phi_{a}^{(1,14,8,8)}\bar{\Phi}_{c}^{(0,8,14,14)}$ & $\Phi_{(0,8,14,14)}^{(0,2,8,8)}\bar{\Phi}_{c}^{(0,8,14,14)}$ & $(2,2,2,2)$ \\
&  & (-2,2) & 6 &  $\Phi_{a}^{(1,10,16,4)}\bar{\Phi}_{c}^{(1,10,16,4)}$ & $\Phi_{c}^{(1,10,16,4)}\bar{\Phi}_{c}^{(1,10,16,4)}$ & $(1,10,16,4)$ \\

 \hline

    1 & $H^{3,3}$ & (-3,3) &  1 &$\Phi^{(1,16,16,16)}_a\bar{\Phi}^{(1,16,16,16)}_c$ & $\Phi^{(1,16,16,16)}_c\bar{\Phi}^{(1,16,16,16)}_c$ & $(1,16,16,16)$ \\

     \hline

 \hline 
 
\end{tabular}
\end{center}
\end{table}


\begin{table}[H]
\caption{$(a,c)$ fields for $(1_A)^1(16_E)^3$ with $\tilde{G}^*=\langle (2,2,2,2),(0,4,4,10),(0,4,10,4)\rangle$   }
\label{tabular:5}
\begin{center}
\begin{tabular}{ |c|c|c|c|c|c|c|} 
 \hline
 Nmb. & $H^{p,q}$ & $(Q_L,Q_R)$ & \# & Field: $\Phi^{\vec{l}}_{a}\bar{\Phi}^{\vec{\bar{l}}}_{c}$ & Obtained from: $\Phi^{\vec{l}}_{\vec{q}}\bar{\Phi}^{\vec{\bar{l}}}_{c}$ & $\vec{w}^*\in \tilde{G}^*$   \\
 \hline
 \hline

     1& $H^{0,0}$ & (0,0) & 1 &  $I\times\bar{I}$ & Untwisted &  \\

     \hline 
23 & $H^{1,1}$& (-1,1) & 1 &  $\Phi_{a}^{(0,6,6,6)}\bar{\Phi}_{c}^{(0,6,6,6)}$ & $\Phi_{c}^{(0,6,6,6)}\bar{\Phi}_{c}^{(0,6,6,6)}$ &(0,6,6,6)  \\
& & (-1,1) & 1 &  $\Phi_{a}^{(1,4,4,4)}\bar{\Phi}_{c}^{(1,4,4,4)}$ & $\Phi_{c}^{(1,4,4,4)}\bar{\Phi}_{c}^{(1,4,4,4)}$ & (1,4,4,4) \\
& & (-1,1) & 3 &  $\Phi_{a}^{(0,4,4,10)}\bar{\Phi}_{c}^{(0,4,4,10)}$ & $\Phi_{c}^{(0,4,4,10)}\bar{\Phi}_{c}^{(0,4,4,10)}$ &(0,4,4,10)  \\
& & (-1,1) & 3 &  $\Phi_{a}^{(1,0,6,6)}\bar{\Phi}_{c}^{(1,0,6,6)}$ &  $\Phi_{c}^{(1,0,6,6)}\bar{\Phi}_{c}^{(1,0,6,6)}$ & (1,0,6,6) \\
& & (-1,1) & 3 &  $\Phi_{a}^{(1,0,0,12)}\bar{\Phi}_{c}^{(1,0,0,12)}$ & $\Phi_{c}^{(1,0,0,12)}\bar{\Phi}_{c}^{(1,0,0,12)}$ & (1,0,0,12) \\
& & (-1,1) & 6 &  $\Phi_{a}^{(0,0,6,12)}\bar{\Phi}_{c}^{(0,0,6,12)}$ & $\Phi_{c}^{(0,0,6,12)}\bar{\Phi}_{c}^{(0,0,6,12)}$ & (0,0,6,12) \\

& & (-1,1) & 3 &  $\Phi_{a}^{(0,8,2,8)}\bar{\Phi}_{c}^{(0,8,8,2)}$ & $\Phi_{(0,8,8,2)}^{(0,8,14,8)}\bar{\Phi}_{c}^{(0,8,8,2)}$ &(0,8,14,14)   \\
& & (-1,1) & 3 &  $\Phi_{a}^{(0,2,8,8)}\bar{\Phi}_{c}^{(0,8,8,2)}$ & $\Phi_{(0,8,8,2)}^{(0,14,8,8)}\bar{\Phi}_{c}^{(0,8,8,2)}$ &(0,14,8,14)   \\

     \hline
    1&  $H^{0,3}$ & (0,3) &  1 & $I\times\bar{\Phi}^{(1,16,16,16)}_c$ & $\Phi^{(1,0,0,0)}_{(1,16,16,16)}\bar{\Phi}^{(1,16,16,16)}_c$ & $(2,8,8,8)$ \\

\hline 

   14 & $H^{1,2}$ & (-1,2) & 1 &  $\Phi^{(0,6,6,6)}_a\bar{\Phi}^{(1,10,10,10)}_c$ & $\Phi^{(1,6,6,6)}_{(1,10,10,10)}\bar{\Phi}^{(1,10,10,10)}_c$ & $(2,8,8,8)$ \\

   & &(-1,2) & 1 &  $\Phi^{(1,4,4,4)}_a\bar{\Phi}^{(0,12,12,12)}_c$ & $\Phi^{(0,4,4,4)}_{(0,12,12,12)}\bar{\Phi}^{(0,12,12,12)}_c$ & $(2,8,8,8)$ \\

   & &(-1,2) & 6 &  $\Phi^{(0,12,6,0)}_a\bar{\Phi}^{(1,4,10,16)}_c$ & $\Phi^{(1,12,6,0)}_{(1,4,10,16)}\bar{\Phi}^{(1,4,10,16)}_c$ & $(2,8,8,8)$ \\

   & &(-1,2) & 3 &  $\Phi^{(0,2,8,8)}_a\bar{\Phi}^{(0,8,14,14)}_c$ & $\Phi^{(0,14,8,8)}_{(0,8,14,14)}\bar{\Phi}^{(0,8,14,14)}_c$ & $(0,14,2,2)$ \\

   & &(-1,2) & 3 &  $\Phi^{(1,2,2,8)}_a\bar{\Phi}^{(1,8,8,14)}_c$ & $\Phi^{(1,14,14,8)}_{(1,8,8,14)}\bar{\Phi}^{(1,8,8,14)}_c$ & $(1,14,14,2)$ \\

   \hline 
  14 & $H^{2,1}$ & (-2,1) & 1 &  $\Phi^{(1,10,10,10)}_a\bar{\Phi}^{(0,6,6,6)}_c$ & $\Phi^{(0,10,10,10)}_{(0,6,6,6)}\bar{\Phi}^{(0,6,6,6)}_c$ & $(2,8,8,8)$ \\

  & &(-2,1) & 1 &  $\Phi^{(0,12,12,12)}_a\bar{\Phi}^{(1,4,4,4)}_c$ & $\Phi^{(1,12,12,12)}_{(1,4,4,4)}\bar{\Phi}^{(1,4,4,4)}_c$ & $(2,8,8,8)$ \\

  & &(-2,1) & 6 &  $\Phi^{(1,4,10,16)}_a\bar{\Phi}^{(0,12,6,0)}_c$ & $\Phi^{(0,4,10,16)}_{(0,12,6,0)}\bar{\Phi}^{(0,12,6,0)}_c$ & $(2,8,8,8)$ \\

  & &(-2,1) & 3 &  $\Phi^{(0,8,14,14)}_a\bar{\Phi}^{(0,2,8,8)}_c$ & $\Phi^{(0,8,2,2)}_{(0,2,8,8)}\bar{\Phi}^{(0,2,8,8)}_c$ & $(0,14,2,2)$ \\

    & &(-2,1) & 3 &  $\Phi^{(1,8,8,14)}_a\bar{\Phi}^{(1,2,2,8)}_c$ & $\Phi^{(1,8,8,2)}_{(1,2,2,8)}\bar{\Phi}^{(1,2,2,8)}_c$ & $(1,14,14,2)$ \\

     \hline
    1& $H^{3,0}$ & (-3,0) & 1 &  $\Phi^{(1,16,16,16)}_a\times\bar{I}$ & $\Phi^{(0,16,16,16)}_{(0,0,0,0)}\times\bar{I}$ & $(2,8,8,8)$ \\
    \hline

23 & $H^{2,2}$ & (-2,2) & 1 &  $\Phi_{a}^{(0,12,12,12)}\bar{\Phi}_{c}^{(0,12,12,12)}$ & $\Phi_{c}^{(0,12,12,12)}\bar{\Phi}_{c}^{(0,12,12,12)}$ &(0,12,12,12)  \\
& & (-2,2) & 1 &  $\Phi_{a}^{(1,10,10,10)}\bar{\Phi}_{c}^{(1,10,10,10)}$ & $\Phi_{c}^{(1,10,10,10)}\bar{\Phi}_{c}^{(1,10,10,10)}$ &(1,10,10,10)  \\

   & &(-2,2) & 3 &  $\Phi^{(0,4,16,16)}_a\bar{\Phi}^{(0,4,16,16)}_c$ & $\Phi^{(0,4,16,16)}_{c}\bar{\Phi}^{(0,4,16,16)}_c$ & $(0,4,16,16)$ \\

   & &(-2,2) & 3 &  $\Phi^{(1,6,12,12)}_a\bar{\Phi}^{(1,6,12,12)}_c$ & $\Phi^{(1,6,12,12)}_{c}\bar{\Phi}^{(1,6,12,12)}_c$ & $(1,6,12,12)$ \\

   & &(-2,2) & 3 &  $\Phi^{(0,16,10,10)}_a\bar{\Phi}^{(0,16,10,10)}_c$ & $\Phi^{(0,16,10,10)}_{c}\bar{\Phi}^{(0,16,10,10)}_c$ & $(0,16,10,10)$ \\

   & &(-2,2) & 6 &  $\Phi^{(1,4,10,16)}_a\bar{\Phi}^{(1,4,10,16)}_c$ & $\Phi^{(1,4,10,16)}_c\bar{\Phi}^{(1,4,10,16)}_c$ & $(1,4,10,16)$ \\
   
   & &(-2,2) & 3 &  $\Phi^{(1,8,14,8)}_a\bar{\Phi}^{(1,8,8,14)}_c$ & $\Phi^{(1,8,2,8)}_{(1,8,8,14)}\bar{\Phi}^{(1,8,8,14)}_c$ & $(1,8,2,2)$ \\

    & &(-2,2) & 3 &  $\Phi^{(1,14,8,8)}_a\bar{\Phi}^{(1,8,8,14)}_c$ & $\Phi^{(1,2,8,8)}_{(1,8,8,14)}\bar{\Phi}^{(1,8,8,14)}_c$ & $(1,2,8,2)$ \\

    \hline

    1 & $H^{3,3}$ & (-3,3) &  1 &$\Phi^{(1,16,16,16)}_a\bar{\Phi}^{(1,16,16,16)}_c$ & $\Phi^{(1,16,16,16)}_c\bar{\Phi}^{(1,16,16,16)}_c$ & $(1,16,16,16)$ \\
    \hline

\end{tabular}
\end{center}
\end{table}


\begin{thebibliography}{99}





\bibitem{BDFM:1988}
T.~Banks, L.~Dixon, D.~Friedan and E.~Martinec, "Phenomenology and conformal field theory or can string theory predict the weak mixing angle?",
\href{https://www.sciencedirect.com/science/article/abs/pii/0550321388905512}
{Nucl. Phys. B \textbf{299}, 613 (1988)}

\bibitem{CHSW:1985}
P.~Candelas, G.T.~Horowitz, A.~Strominger and E.~Witten, "Vacuum configurations for superstrings", 
\href{https://www.sciencedirect.com/science/article/abs/pii/0550321385906029}
{Nucl. Phys. B \textbf{258}, 46 (1985)}


\bibitem{Gepner:1987vz}
D.~Gepner,
``Exactly Solvable String Compactifications on Manifolds of SU(N) Holonomy,''
\href{https://www.sciencedirect.com/science/article/abs/pii/0370269387909385}{Phys. Lett. B \textbf{199}, 380-388 (1987)}

\bibitem{Gepner:1988}
D. ~Gepner, "Space-time supersymmetry in compactified string theory and superconformal models",
\href{https://www.sciencedirect.com/science/article/abs/pii/0550321388903975}
{Nucl. Phys. B \textbf{296}, 757 (1988)}

\bibitem{BP}
A.A.~Belavin and S.E.~Parkhomenko, "Explicit construction of N=2 Superconformal Orbifolds", \href{https://link.springer.com/article/10.1134/S0040577921100044}{Theor. Math. Phys. \textbf{209:1}, 1367-1386 (2021)}

\bibitem{BBP}
A.~Belavin, V.~Belavin and S.~Parkhomenko, "Explicit construction of N=2 SCFT orbifold models. Spectral flow and mutual locality",
\href{https://www.sciencedirect.com/science/article/pii/S0550321322002425?via%3Dihub}{Nucl. Phys. B \textbf{982}, 115891 (2022)}
\href{https://arxiv.org/abs/2206.03472}{[arXiv:2206.03472 [hep-th]]}

\bibitem{IIAB}
A.~Belavin and S.~Parkhomenko, "Mirror symmetry and new approach to constructiong orbifolds of Gepner models", 
\href{https://www.sciencedirect.com/science/article/pii/S0550321323003589}
{Nucl. Phys. B \textbf{998}, 116431 (2024)}
\href{https://arxiv.org/abs/2311.15403}
{ArXiv:hep-th/2311.15403}

\bibitem{Cappelli:1987xt}
A.~Cappelli, C.~Itzykson and J.~B.~Zuber,
``The ADE Classification of Minimal and A1(1) Conformal Invariant Theories,''
\href{https://link.springer.com/article/10.1007/BF01221394}{Commun. Math. Phys. \textbf{113}, 1 (1987)}

\bibitem{Lerche:1989uy}
W.~Lerche, C.~Vafa and N.~P.~Warner,
``Chiral Rings in N=2 Superconformal Theories,''
\href{https://www.sciencedirect.com/science/article/abs/pii/0550321389904744}{Nucl. Phys. B \textbf{324}, 427-474 (1989)}


\bibitem{Gepner:Parafermion}
D.~Gepner and Z.~a.~Qiu,
``Modular Invariant Partition Functions for Parafermionic Field Theories,''
\href{https://www.sciencedirect.com/science/article/abs/pii/0550321387903488?via%3Dihub}{Nucl. Phys. B \textbf{285} (1987), 423}

\bibitem{Gepner:Lectures}
D.~Gepner,
``Lectures on N=2 String theory,''
PUPT-1121 (1989).

\bibitem{Gannon:1999cp}
T.~Gannon,
``The Cappelli-Itzykson-Zuber A-D-E classification,''
\href{https://www.worldscientific.com/doi/abs/10.1142/S0129055X00000265?srsltid=AfmBOop_6yULtRD6jmk5aWtU0dO03ANPZOUSNKmB0yRMD9JXcUcgfqJK}{Rev. Math. Phys. \textbf{12}, 739-748 (2000)}, \href{https://arxiv.org/abs/math/9902064}{[arXiv:math/9902064]}

\bibitem{Gannon:1996hp}
T.~Gannon,
``U(1)-m modular invariants, N=2 minimal models, and the quantum Hall effect,''
\href{https://www.sciencedirect.com/science/article/abs/pii/S0550321397000321}{Nucl. Phys. B \textbf{491}, 659-688 (1997)}
\href{https://arxiv.org/abs/hep-th/9608063}{[arXiv:hep-th/9608063 [hep-th]]}.

\bibitem{Feigin}
B.~L.~Feigin and A.~M.~Semikhatov,
``Free field resolutions of the unitary N=2 superVirasoro representations,''
\href{https://arxiv.org/abs/hep-th/9810059}{[arXiv:hep-th/9810059 [hep-th]]}.

\bibitem{FSST}B. L. Feigin, A. M. Semikhatov, V. A. Sirota and I. Yu. Tipunin, "Resolutions and characters of irreducible representations of the N=2 superconformal algebra",
\href{https://arxiv.org/abs/hep-th/9805179}{[arXiv:hep-th/9805179v3 [hep-th]]}.




\bibitem{Fuchs:1991vu}
J.~Fuchs, A.~Klemm and M.~G.~Schmidt,
``Orbifolds by cyclic permutations in Gepner type superstrings and in the corresponding Calabi-Yau manifolds,''
\href{https://www.sciencedirect.com/science/article/abs/pii/S0003491605800016}{Annals Phys. \textbf{214}, 221-257 (1992)}

\bibitem{Fuchs:1989pt}
J.~Fuchs, A.~Klemm, C.~Scheich and M.~G.~Schmidt,
``Gepner Models With Arbitrary Affine Invariants and the Associated Calabi-yau Spaces,''
\href{https://www.sciencedirect.com/science/article/abs/pii/0370269389907508}{Phys. Lett. B \textbf{232}, 317-322 (1989)}

\bibitem{Lynker}
M.~Lynker and R.~Schimmrigk,
``ADE quantum Calabi-Yau manifolds,''
\href{https://www.sciencedirect.com/science/article/abs/pii/055032139090536M?via%3Dihub}{Nucl. Phys. B \textbf{339} (1990), 121-157}


\bibitem{BH} P. Berglund and T. Hubsch, "A generalized construction of mirror manifolds",
AMS/IP Stud.
\href{https://www.sciencedirect.com/science/article/abs/pii/055032139390250S?via%3Dihub}{Nucl. Phys. B \textbf{393} (1993), 377-391}
\href{https://arxiv.org/abs/hep-th/9201014}{arXiv:hep-th/9201014}

\bibitem{Kra} M.Krawitz, "FJRW rings and Landau-Ginsburg Mirror Symmetry", 
\href{https://arxiv.org/abs/0906.0796}{arXiv:math.AG/0906.0796}

\bibitem{EOTY}T.Eguchi, H.Ooguri, A.Taormina and S-K.Yang,
"Superconformal algebras and string compactification on manifolds with $SU(n)$ holonomy",
\href{https://www.sciencedirect.com/science/article/abs/pii/0550321389904549}{ Nucl. Phys. B \textbf{315}, 193 (1989)}

\bibitem{Parkhomenko:2022kju}
S.~Parkhomenko,
``Spectral flow construction of mirror pairs of CY orbifolds,''
\href{https://www.sciencedirect.com/science/article/pii/S055032132200356X}{Nucl. Phys. B \textbf{985}, 116005 (2022)}
\href{https://arxiv.org/abs/2208.11612}{[arXiv:2208.11612 [hep-th]]}

\bibitem{ShYan} A.N.Schellekens and S.Yankielowicz, ``Exceptional Modular Invariants of $N=2$ Tensor Products,''
\href{https://www.sciencedirect.com/science/article/abs/pii/037026939091592Y}
{Phys. Lett. B \textbf{242}, 45 (1990)}

\bibitem{SS}A. Schwimmer and N. Seiberg, "Comments on the N=2, N=3, N=4 Superconformal Algebras in Two-Dimensions", 
\href{https://www.sciencedirect.com/science/article/abs/pii/0370269387905661?via%3Dihub}{Phys. Lett. B \textbf{184}, 191 (1987)}


\bibitem{FST}B. L. Feigin, A. M. Semikhatov and I. Yu. Tipunin, ``A Semiinfinite construction of unitary N=2 modules,''
\href{https://arxiv.org/abs/hep-th/0004066}{[arXiv:hep-th/0004066]}.

\bibitem{ZF}A.B. Zamolodchikov and V.A. Fateev, "Disorder fields in two-dimensional conformal quantum field theory and $N=2$ extended supersymmetry",
JETP, 1986, Vol. 63, No. 5, p. 913




\bibitem{Parkhomenko:2024mxq}
S.~Parkhomenko,
``Conformal bootstrap and Mirror symmetry of states in Gepner models,''
\href{https://link.springer.com/article/10.1007/JHEP11(2024)104}{JHEP \textbf{11} (2024), 104}\href{https://arxiv.org/abs/2407.07555}{[arXiv:2407.07555 [hep-th]]}

\bibitem{ABBE}
A.~Belavin and B.~Eremin,
``On the equivalence of Batyrev and BHK mirror symmetry constructions,''
\href{https://www.sciencedirect.com/science/article/pii/S0550321320303564?via%3Dihub}{Nucl. Phys. B \textbf{961}, 115271 (2020)}, \href{https://arxiv.org/abs/2010.07687}{[arXiv:2010.07687 [hep-th]]}


\bibitem{Schimmrigk}
R.~Schimmrigk,
``A New Construction of a Three Generation Calabi-yau Manifold,''
\href{https://www.sciencedirect.com/science/article/abs/pii/0370269387912184?via%3Dihub}{Phys. Lett. B \textbf{193} (1987), 175}

\bibitem{Gepner:1987hi}
D.~Gepner,
``String Theory on Calabi-Yau Manifolds: The Three Generations Case,''
\href{https://arxiv.org/abs/hep-th/9301089}{[arXiv:hep-th/9301089]}

\bibitem{GrPl}
B.R. Greene and M.R. Plesser, "Duality in Calabi-Yau moduli space",
\href{https://www.sciencedirect.com/science/article/abs/pii/055032139090622K}{Nucl. Phys. B \textbf {338} (1990), 15}






































\end{thebibliography}
\end{document}